\documentclass[manuscript]{acmart}

\AtBeginDocument{%
  }

\acmJournal{TOSEM}

\usepackage{array}
\usepackage{textcomp}
\usepackage{stfloats}
\usepackage{url}
\usepackage{verbatim}
\usepackage{graphicx}
\usepackage{xcolor}
\usepackage[table]{xcolor}
\usepackage[flushleft]{threeparttable}
\usepackage{ragged2e}
\usepackage{tabularx}
\usepackage{multirow}
\usepackage{multicol}
\usepackage{fontawesome5}
\usepackage[many]{tcolorbox} 

\newcolumntype{P}[1]{>{\RaggedRight\hspace{0pt}}p{#1}}
\definecolor{tableGray}{RGB}{243, 244, 245}
\definecolor{borderGray}{RGB}{229, 230, 233}

\newtcolorbox{boxA}{
    colback = tableGray, 
    boxrule = 0pt  
}

\def\blackbar#1#2{
   {\color{black}\rule{#1mm}{4pt}}  #2}

\def\multicolorbar#1#2#3#4#5#6{
   {\color{black!20}\rule{#1mm}{6pt}}%
   {\color{black!40}\rule{#2mm}{6pt}}%
   {\color{black!60}\rule{#3mm}{6pt}}%
   {\color{black!75}\rule{#4mm}{6pt}}%
   {\color{black!100}\rule{#5mm}{6pt} #6}%
}

\begin{document}

\title[A Theory of the Role of Empathy in SE]{The Role of Empathy in Software Engineering - A Socio-Technical Grounded Theory}

\author{Hashini Gunatilake}
\email{hashini.gunatilake@monash.edu}
\orcid{0000-0002-4823-0214}

\affiliation{%
  \institution{Faculty of Information Technology, Monash University}
  \city{Melbourne}
  \country{Australia}
}

\author{John Grundy}
\orcid{0000-0003-4928-7076}
\affiliation{%
  \institution{Faculty of Information Technology, Monash University}
  \city{Melbourne}
  \country{Australia}}
\email{john.grundy@monash.edu}

\author{Rashina Hoda}
\orcid{0000-0001-5147-8096}
\affiliation{%
  \institution{Faculty of Information Technology, Monash University}
  \city{Melbourne}
  \country{Australia}
}
\email{rashina.hoda@monash.edu}

\author{Ingo Mueller}
\orcid{0000-0003-2240-712X}
\affiliation{%
  \institution{Monash Centre for Health Research \& Implementation, Monash University \& Monash Health}
  \city{Melbourne}
  \country{Australia}}
\email{ingo.mueller@monash.edu}

\renewcommand{\shortauthors}{Gunatilake et al.}

\begin{abstract}
Empathy, defined as the ability to understand and share others' perspectives and emotions, is essential in software engineering (SE), where developers often collaborate with diverse stakeholders. It is also considered as a vital competency in many professional fields such as medicine, healthcare, nursing, animal science, education, marketing, and project management. Despite its importance, empathy remains under-researched in SE. To further explore this, we conducted a socio-technical grounded theory (STGT) study through in-depth semi-structured interviews with 22 software developers and stakeholders. \textcolor{black}{Our study explored the role of empathy in SE and how SE activities and processes can be improved by considering empathy. Through applying the systematic steps of STGT data analysis and theory development, we developed a theory that explains the role of empathy in SE. 
Our theory details the contexts in which empathy arises, the conditions that shape it, the causes and consequences of its presence and absence.} We also identified contingencies for enhancing empathy or overcoming barriers to its expression. Our findings provide practical implications for SE practitioners and researchers, offering a deeper understanding of how to effectively integrate empathy into SE processes.
\end{abstract}



\received{03 December 2024}
\received[revised]{1 April 2025}
\received[revised]{4 July 2025}
\received[accepted]{26 August 2025}

\maketitle

\section{Introduction} \label{sec:Introduction}

Software engineering (SE) is a highly technical and deeply collaborative field, where the success of projects often depends on effective interactions between developers and various stakeholders, including product owners, project managers, testers, and customers \cite{ojastu2014expectations, alami2016projectfail}. These interactions significantly influence the design, development, and delivery of software. These interactions are not purely technical but are deeply influenced by a range of human aspects \cite{gunatilake2024SLR}, including empathy, a factor increasingly recognised as pivotal in collaborative work environments \cite{gunatilake2023empathy}. We define empathy as \textit{``the ability to experience the affective and cognitive states of another person, while maintaining a distinct self, in order to understand the other''} \cite{guthridge2021taxonomy}. This meta-definition, derived through an inductive conceptual content analysis of existing empathy definitions, offers a clear framework for understanding its fundamental dimensions. 
\textcolor{black}{In the context of SE, empathy improves software quality, resulting in fewer bugs and better alignment with user needs \cite{cerqueira2023thematic}. It fosters trust, reduces conflicts, and enhances communication among stakeholders, contributing to a healthier work environment. Empathy also supports practitioners' well-being and reduces stress and anxiety \cite{cerqueira2024empathy}.}
Despite its importance, the role of empathy in SE has been largely overlooked in SE research to date \cite{gunatilake2024SLR, gunatilake2024enablers, gunatilake2023empathy}. 

The under-representation of empathy in SE literature points to a significant gap in understanding the role of empathy in developer-stakeholder interactions and how it impacts the effectiveness of these interactions. While the effects of empathy have been extensively studied in other disciplines  \textcolor{black}{such as healthcare \cite{hojat2016empathy, decety2015empathy, derksen2013effectiveness, decety2020empathy, moudatsou2020role, de2022impact}, engineering \cite{hess2016voices, strobel2013empathy}, and sales \cite{aggarwal2005salesperson}}, only \textcolor{black}{three} studies have specifically examined its effects in SE \textcolor{black}{\cite{cerqueira2024empathy, cerqueira2023thematic, lind2024empathy}}. Both these studies provide valuable insights by analysing grey literature, particularly the blog posts from developers,\footnote{https://dev.to} offering an important glimpse into how empathy is perceived within the developer community.
However, while these reflections offer a useful starting point, they do not capture the full depth and complexity of the lived experiences in day-to-day developer-stakeholder interactions. Blog posts, while offering direct insights, often capture more surface-level experiences and may not fully convey the richer, more nuanced perspectives that emerge from in-depth conversations with practitioners hearing their first-hand experiences. Direct engagement allows probing into specific details, clarifying ambiguities, and understanding the nuances of real-time decision making and interpersonal dynamics, providing a depth that blog posts alone cannot achieve. Recognising this gap, our study aims to build on the existing work by providing a more in-depth empirical investigation through direct interaction with practitioners. This allows us to explore how empathy manifests in real-world software development contexts and how it influences developer-stakeholder relationships, providing a more comprehensive understanding of its impact.

Our previous work \cite{gunatilake2024enablers} identified key enablers and barriers to empathy in interactions between developers and end-users, offering insights into causes of empathy and its absence, and strategies to overcome barriers to empathy. However, it did not provide a holistic understanding of the role of empathy in SE. Existing studies \textcolor{black}{\cite{gunatilake2023empathy, cerqueira2023thematic, cerqueira2024empathy, lind2024empathy}} have examined specific aspects of empathy in SE, but no study has systematically explored its broader role across SE practices. To address this gap, we formulated below research questions (RQs) to guide an exploratory investigation into the role of empathy in SE:
\begin{enumerate}
    \item [\textbf{RQ1:}]\textbf{What is the role of empathy in SE, and how does it impact software developers and their stakeholders?} Given the limited understanding of empathy in SE, this RQ aims to provide a holistic perspective by exploring the context in which empathy occurs, the conditions that give rise to it, the causes and consequences of both empathy and its absence, and various factors that influence the empathy of developers and stakeholders.


        
        
        
        

    \item [\textbf{RQ2:}]\textbf{How can the SE activities and processes be improved considering empathy of software developers and their stakeholders?} This RQ examines the strategies employed by developers and stakeholders to foster empathy and address empathy-related challenges in SE.
\end{enumerate}


\textcolor{black}{To explore these RQs, we conducted a Socio-Technical Grounded Theory (STGT) study involving 22 interviews with developers and stakeholders, analysed using the basic and advanced stages of STGT for data analysis and theory development \cite{hoda2022STGT, hoda2024qualitative}. This approach allowed us to explore the nuanced ways in which empathy shapes SE processes, influencing both technical and social dimensions of SE. 
Our RQs guided the study by clarifying its focus and refining its scope \cite{hoda2024qualitative}. Applying STGT, we adopted an inductive approach in the basic stage to analyse participant experiences and construct an initial conceptual understanding. Encouraged by the emergence of a theoretical structure, we proceeded to the advanced stage of STGT, refining and integrating the emerging concepts into a more comprehensive theory on empathy in SE. Our theory on \textit{the role of empathy in developer-stakeholder interactions within SE} explains the \textit{contexts} in which empathy arises, the \textit{conditions} that shape it, the \textit{causes} and \textit{consequences} of its presence and absence, and \textit{contingencies} for enhancing empathy or overcoming barriers to its expression.}
The key contributions of this work include the identification of: 
\begin{itemize}
    \item A novel theory that explains the role of empathy in SE, including: 

    \begin{itemize}
        \item \textcolor{black}{The causes of empathy and lack of empathy};
        \item \textcolor{black}{The consequences of empathy and lack of empathy, including their direct impact on technical outcomes in SE as well as the mental health and well-being of software practitioners;}
        \item \textcolor{black}{The strategies used to enhance empathy and mitigate empathy barriers};
        \item \textcolor{black}{The factors that influence empathy};
    \end{itemize}
    \item \textcolor{black}{Key future research directions for SE researchers;}
    \item \textcolor{black}{Practical implications for software practitioners, highlighting how empathy can be supported and leveraged to improve team collaboration, development processes, and practitioner well-being.}
\end{itemize}


\subsection*{Glossary of Terms} \label{sec:Glossary}
We use a range of terminology throughout our paper and brief definitions of these terms are provided in Table \ref{tab:Glossary of Terms}, sorted in alphabetical order for clarity.

\begin{table*}[ht]
\setlength{\belowrulesep}{0pt}
\setlength{\extrarowheight}{.5ex}
    
\begin{threeparttable}
    \footnotesize
    \centering
    \caption{Glossary of Terms}
    \label{tab:Glossary of Terms}
    \begin{tabular}{P{0.19\linewidth} P{0.75\linewidth}}
        \toprule
         \textbf{Term} & \textbf{Definition}  \\
         \midrule
         


         \rowcolor{tableGray}
         Affective/Emotional Emp. & The ability of a person to perceive and share other individual's emotional states and feelings \cite{de2008putting, ilgunaite2017measuring}.\\

        Behavioural Empathy & Refers to the outward expression of empathy through actions and behaviours \cite{clark2019feel}.\\ 

         \rowcolor{tableGray}
         Causes & The reasons that lead to the occurrence of the central phenomenon.\\
         
         Central Phenomenon & The focal point of the study.\\

         \rowcolor{tableGray}
         Cognitive Empathy & The ability of a person to consciously detect and understand the internal states of others \cite{goldman2011two, wallmark2018neurophysiological}.\\

         Collectivism & Societies in which people from birth onwards are integrated into strong, cohesive in-groups, which throughout people's lifetime continue to protect them in exchange for unquestioning loyalty \cite{hofstede1991empirical}.\\
     
         \rowcolor{tableGray}
         Conditions & The factors that are prerequisites for the central phenomenon to manifest and alter it in some way.\\ 
         Consequences & Describe the effects of the central phenomenon.\\
         
         \rowcolor{tableGray}
         Context & The setting in which the central phenomenon occurs.\\
         Contingencies & The strategies used to address the central phenomenon.\\
         
         \rowcolor{tableGray}
         Covariances & Capture the relationships where changes in one category imply changes in another.\\


         Empathic Concern & The tendency of a person to experience feelings of warmth, compassion and concern for others undergoing negative experiences \cite{davis1980IRI}.\\
         
         \rowcolor{tableGray}
         Empathy & The ability to experience affective and cognitive states of another person, while maintaining a distinct self, in order to understand the other \cite{guthridge2021taxonomy}.\\

         Individualism & Societies in which the ties between individuals are loose; everyone is expected to look after themself \& their immediate family \cite{hofstede1991empirical}.\\



    
         \rowcolor{tableGray}
         Personal Distress & The tendency to experience distress and discomfort in response to extreme distress in others \cite{davis1980IRI}.\\

         Perspective Taking & The ability of a person to see the situation from another person's perspective \cite{reniers2011QCAE, ilgunaite2017measuring}.\\

         
         \rowcolor{tableGray}
         Software Developer & \textcolor{black}{Individuals primarily responsible for developing and maintaining software, including tasks such as coding new features, fixing issues, and ensuring the ongoing functionality and performance of the software.}\\
         
         Software Stakeholder & Individuals or groups with vested interests in project outcome. In this study, we focus on stakeholders whose primary responsibilities involve providing requirements, feedback, and expectations to developers.\\


    
         
         \bottomrule
    \end{tabular}
    \end{threeparttable}
\end{table*}

\section{Related Work} \label{sec:Related Works} 

Empathy is generally recognised as having three dimensions: \textit{cognitive}, \textit{affective}, and \textit{behavioural} empathy \cite{clark2019feel, cuff2016empathy}. However, debate persists on whether empathy is primarily cognitive, affective, or a combination of both \cite{clark2019feel, hojat2007empathy}. This distinction is important for understanding how empathy can be conceptualised and applied in software development, where interactions are often technical and task-focused. Empathy has been studied from multiple perspectives, including evolutionary, psychological, neuroscientific, moral, political, economic, and cultural viewpoints, leading to diverse conceptualisations and a lack of a unified definition \cite{guthridge2020critical}. Guthridge and Giummarra identified six key dimensions of empathy \cite{guthridge2021taxonomy}, including cognitive and affective states, supporting the widely accepted understanding that empathy involves both cognitive and affective components \cite{decety2011neuroevolution}.
Recent studies have explored \textit{how empathy is conceptualised within SE}, identifying key themes such as understanding, compassion, perspective-taking, embodiment, and emotional sharing which are grouped under cognitive, emotional, and compassionate empathy \cite{cerqueira2024empathy, cerqueira2023thematic}. These insights provide a foundation for understanding how empathy shapes SE practices, including team interactions and stakeholder communication. 

While research on causes, consequences, and influencing factors of empathy remain limited within SE, these aspects have been extensively explored in other domains. 
\textit{The causes of empathy} have been widely studied across disciplines \textcolor{black}{such as} psychology \cite{batson1996prior, eklund2009similar, heinke2009cultural, yaghoubi2023young}, neuroscience \cite{preis2012pain, xu2009pain, yaghoubi2021histories}, and philosophy \cite{snow2000empathy}. Research consistently highlights similarities, such as shared experiences or demographic traits \cite{jami2023interaction}, as key factors that enhance empathy by fostering connection. Familiarity is another cause of empathy \cite{motomura2015interaction, airenti2015cognitive}, while cultural backgrounds play a significant role in shaping empathic behaviour \cite{jami2023interaction}. 
Research on the \textit{causes of lack of empathy} has mainly focused on healthcare, identifying several key factors that hinder empathy \cite{derksen2016managing, howick2017barriers, halpern2003clinical}. These include time pressure, conflicting priorities, and bureaucracy \cite{howick2017barriers}. Among physicians, barriers such as anxiety, difficulty of understanding patients' emotional needs, and negative emotions from tensions with patients contributed to a lack of empathy \cite{halpern2003clinical}. For oncology nurses, empathy is challenged by lack of compassion, job strain, prioritising tasks over patient care, and nurse-patient gender imbalance \cite{taleghani2018barriers}. In cancer care, maintaining empathy is difficult due to feelings of futility. 
Overall, factors \textcolor{black}{such as} familiarity, similarities, and culture fostered empathy, while time pressure, negative emotions, and task-centeredness hindered it.

\textit{\textcolor{black}{Consequences of empathy}} is another overlooked area in SE research \cite{cerqueira2023thematic, cerqueira2024empathy, lind2024empathy}. 
However, it is highly researched in healthcare \cite{hojat2016empathy, decety2015empathy, harscher2018impact, derksen2013effectiveness, decety2020empathy, chen2024impact, moudatsou2020role, de2022impact}, and we also found studies in other areas such as engineering \cite{hess2016voices, strobel2013empathy}, social care \cite{moudatsou2020role}, intergroup relations \cite{stephan1999role}, sales \cite{aggarwal2005salesperson}, business leadership \cite{rahman2013impact}, hospitality management \cite{min2015factors}, and social entrepreneurial intentions \cite{usman2022impact}. 
Studies in SE show that empathy enhances software quality, fosters trust, reduces conflicts, and improves communication among stakeholders \cite{cerqueira2023thematic, cerqueira2024empathy, lind2024empathy}. It also promotes well-being, teamwork, and collaboration, addressing burnout and improving the work environment. In healthcare, empathy improves patient outcomes, physician well-being, patient satisfaction, and treatment adherence, while reducing professional stress and improving trust between physicians and patients \cite{hojat2016empathy, decety2015empathy, derksen2013effectiveness, decety2020empathy, moudatsou2020role, de2022impact}. In medical education, higher empathy scores among students are linked to better mental health, communication skills, and interest in people-oriented specialities \cite{chen2024impact}. Empathy in engineering enhances design quality, collaboration, and client relationships, fostering mutual trust and respect \cite{hess2016voices, strobel2013empathy}. In social care, empathy builds trust, fosters therapeutic change, and facilitates social change \cite{moudatsou2020role}. It also improves intergroup relations by promoting understanding of different world-views, cultural practices, and emotional connections \cite{stephan1999role}. In sales, empathy strengthens trust and customer satisfaction \cite{aggarwal2005salesperson}, and in business leadership, it improves effectiveness and adaptability \cite{rahman2013impact}. In the hospitality industry, empathy boosts customer satisfaction by making complaint responses more personal \cite{min2015factors}. In social entrepreneurship, empathy motivates individuals to address societal issues, driving positive change \cite{usman2022impact}. These studies highlight the broad impact of empathy across various fields.
Research has also revealed potential \textit{side effects of empathy \textcolor{black}{(negative impact of presence of empathy)}}, challenging the assumption that it is always morally positive \cite{breithaupt2018bad, riess2015impact}. 
\textcolor{black}{A study showed that empathy can lead to emotional distress of clinicians rather than empathic concern, contributing to emotional exhaustion, burnout, and even suicide due to high workloads and overwhelming responsibilities \cite{riess2015impact}. Empathy can also cause personal satisfaction for \textcolor{black}{the} empathiser, even when it harms the target \cite{breithaupt2018bad}.}

Research has identified \textit{factors influencing empathy \textcolor{black}{(conditions)}}, particularly in social work and healthcare settings \cite{moudatsou2020role, van2014psychosocial}. Age, self-reflection, and emotional expression were found to affect empathy levels of female social workers and those with more experience showing higher empathy scores \cite{stanley2020predictors}. Interestingly, female physicians rated their empathic concern higher, but male physicians exhibited more vocal synchrony with patients, and their verbal empathy statements led to higher patient satisfaction \cite{van2014psychosocial}. It is unclear what factors influence empathy in SE contexts.

\textcolor{black}{Empathy has been explored through various \textit{theories and models} across disciplines. However, to the best of our knowledge, no existing theories specifically address empathy within the context of SE. Theodor Lipps’ theory of empathy highlights emotional projection, suggesting that individuals internally imitate others' emotions and project their own feelings onto external stimuli, particularly in aesthetic experiences \cite{burns2021theodor}. His work laid the foundation for later research on mirror neurons, which support the neural basis of empathy.
Hoffman’s theory of moral development focuses on empathic distress, describing how individuals experience distress when witnessing others’ suffering \cite{hoffman1996empathy, wondra2015appraisal}. He outlines five mechanisms that facilitate empathy development, explaining how people either automatically or deliberately relate to others' emotions. 
The appraisal theory of empathy suggests that empathy arises from how we evaluate others' situations, similar to how our own emotions are shaped by our appraisals of events \cite{wondra2015appraisal}. Empathy occurs when an observer appraises a situation similarly to the target, otherwise, a different emotional response is triggered.
Similarly the perception-action model of empathy describes empathy as a shared emotional experience that occurs when one person feels a similar emotion to another by perceiving the other’s emotional state \cite{preston2002empathy, gunatilake2023empathy}. 
The empathy spectrum offers a holistic view of empathy in nursing, identifying four forms: empathy as an incident, a way of knowing, a process, and a way of being \cite{wiseman2007toward, gunatilake2023empathy}. A model of empathy in engineering frames empathy as a teachable skill, a practice orientation, and a professional way of being \cite{walther2017model, gunatilake2023empathy}. It emphasises the interdependence of these dimensions. The skill dimension includes socio-cognitive processes that support communication and decision-making, the orientation dimension reflects mental traits that shape professional engagement, and the being dimension integrates empathy within broader professional values.}

In summary, empathy has gained increasing attention across various disciplines. However, its role in SE remains under-explored, especially regarding its impact on the technical and interdisciplinary dynamics within software development teams. Building on previous work on empathy \cite{gunatilake2023empathy, gunatilake2024enablers}, this study aims to address this gap by conducting an empirical investigation into the real-world practice of empathy in SE.

\section{Research Method} \label{sec:Research Method}

\textcolor{black}{The aim of study is to explore how empathy influences interactions between developers and their stakeholders. To explore this in real-world settings, we decided to conduct semi-structured, in-depth interviews with two primary groups: \textit{software stakeholders} who work with developers and either directly interact with them or provide requirements or feedback (e.g., product owner, tester etc.), and \textit{developers} who directly interact with these stakeholders in gathering requirements or receiving feedback on their software solutions. Given the close involvement of both these groups in the software development process, we recognised these participants would offer practical insights into its intricacies.}

\subsection{\textcolor{black}{Selecting an Empirical Research Method}}
We selected STGT as our empirical research method because our study aligns closely with its socio-technical (ST) research framework \cite{hoda2022STGT, hoda2024qualitative}, across its four dimensions: 

\begin{itemize}
    \item \textcolor{black}{\textit{ST phenomenon:} The role of empathy in SE is inherently a ST phenomenon}, where the social and technical aspects are closely intertwined and need to be comprehensively considered, 
    \item \textcolor{black}{\textit{ST domain and actors:}} The domain is SE, an inherently ST domain and the actors we planned to study are software developers and stakeholders,
    \item \textcolor{black}{\textit{ST researchers:}} Our research team is comprised of experienced ST researchers,
    \item \textcolor{black}{\textit{ST data tools and techniques:}} We used ST tools, techniques and data including NVivo and Zoom, along with data collection methods \textcolor{black}{such as} interviews and questionnaires, to gain deep insights into participants' experiences.
\end{itemize}

\textcolor{black}{In line with the STGT guidelines, we formulated RQs to guide our data collection and analysis. While RQs are not obligatory in STGT studies, they serve a valuable role in focusing the research and clarifying its scope \cite{hoda2024qualitative}. The RQs provided structure to our investigation and clarified which aspects of the ST phenomenon were within scope. They not only guided our efforts but also ensured that the research remained aligned with the participants' experiences. As recommended, we intentionally kept our RQs broad to allow for a comprehensive exploration of the role of empathy.}

\textcolor{black}{In an STGT study, the choice of \textit{research paradigm} is crucial as it influences interviewing, concept development, and the theory formation approach. Traditional \textcolor{black}{grounded theory} methods align with specific paradigms, Glaserian with positivism \cite{glaser2017discovery}, Strauss-Corbinian with symbolic interactionism \cite{strauss1990gt, corbin2014gt} and Charmaz's with constructivism \cite{charmaz2014gt}. In contrast, STGT offers flexibility, allowing researchers to choose the paradigm that best fits their study, provided it is clearly articulated and consistently applied \cite{hoda2022STGT}. Given the interviewer's (first author) industry experience in software development and close interactions with both technical and business stakeholders, a subjective, context-specific, constructivist paradigm was adopted. This paradigm acknowledges the researcher's active role in shaping the interview process, interpreting responses, and developing concepts and categories based on these interactions.}


STGT involves a two-stage approach for data analysis: a \textit{basic stage} for data collection and analysis, and an \textit{advanced stage} for theory development. It offers the flexibility to use only the basic stage or both stages, depending on whether the study involves theory development. \textcolor{black}{The decision to transition to the advanced stage is informed by researcher's familiarity and confidence with data analysis through the basic stage \cite{hoda2024qualitative}. In our study, the co-authors were encouraged by the emergence of a theoretical structure after basic data collection and analysis, and decided to proceed to the advanced stage of theory development.} This led to the development of a theory about the role of empathy in developer - stakeholder interactions. \textcolor{black}{The basic stage is detailed in Section \ref{sec:Basic Data Collection and Analysis}, and the advanced stage is detailed in Section \ref{sec:Theory Development}}. 


\subsection{\textcolor{black}{Basic Stage of Data Collection and Analysis}} \label{sec:Basic Data Collection and Analysis}

\subsubsection{Interview Study Design} \label{sec:Interview Study Design}
\begin{figure}[htbp]
    \centering
    \includegraphics[width=\textwidth]{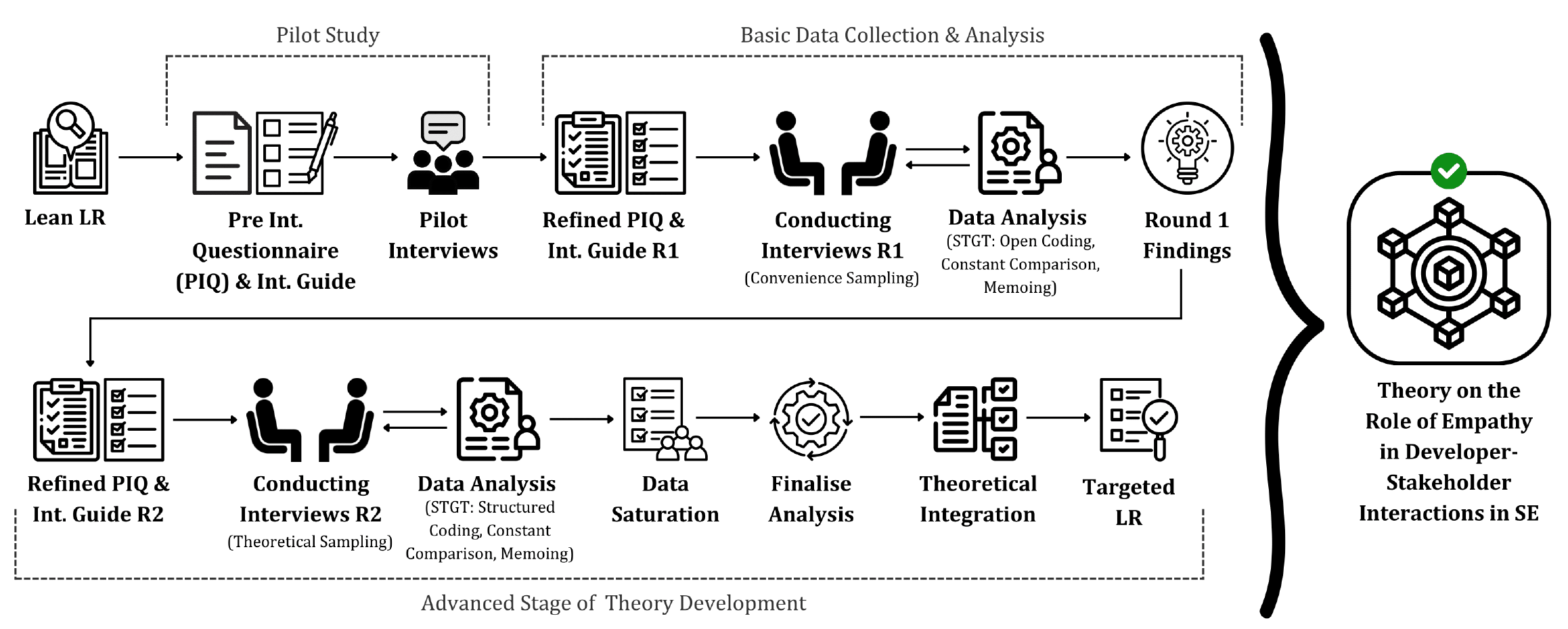}
    \Description{A diagram showing the methodology of the study}
    \caption{Research methodology: Applying basic \& advanced stages of STGT \cite{hoda2024qualitative} to develop the \textit{theory on the role of empathy in developer-stakeholder interactions in SE}. *LR: Literature Review, Int: Interview, R1: Round 1, R2: Round 2, \textcolor{black}{PIQ: Pre Interview Questionnaire}.}
    \label{fig:Methodology}
\end{figure}

In the basic stage, we conducted a \textit{lean literature review} which is a lightweight review, to identify basic research gaps and relevant definitions following the STGT guidelines \cite{hoda2022STGT}. \textcolor{black}{Findings of this review are presented in Section \ref{sec:Introduction} and \ref{sec:Related Works}}. 
The focus of our study\footnote{Approved by Monash Human Research Ethics Committee. ERM Reference Number: 41060} was \textcolor{black}{to explore the role of empathy on the interactions between developers and stakeholders (RQ1) and identify how SE activities and processes can be improved considering their empathy (RQ2).} We developed a pre-interview questionnaire comprising of two main types of questions related to: basic demographics and work experience. The pre-interview questionnaire allowed us to gain a better understanding of the participants and their experience \textcolor{black}{(see Section \ref{sec:Context} for participant details)}, enabling us to focus more effectively on their interactions and impact of empathy, during the interviews.
\textcolor{black}{We created two interview guides for developers and stakeholders comprised of core questions related to the impact and causes of empathy and its absence, and strategies used to enhance empathy or mitigate empathy barriers (see our online Appendix\footnote{\label{onlineappendix}https://github.com/Hashini-G/SupplementaryInfoPackage-6CsStudy}), with the exploration of the impact and causes of empathy aligning with RQ1 and the strategies linked to RQ2.} To uncover nuanced insights, we included situational questions, asking participants to describe scenarios where empathy was shown towards them or not, and where they showed empathy or did not. After creating the interview guides, the first author sought input from more experienced co-authors to further ensure the clarity and appropriateness of questions. 
We conducted a pilot study with a developer and a stakeholder, which helped us to assess the clarity of the questions and gather their suggestions for improving our study. Based on their feedback, we further modified our interview questions and flow of the interview to enhance the clarity of the study. Figure \ref{fig:Methodology} shows an overview of our study methodology.

\subsubsection{Basic Data Collection} \label{sec:Data Collection}

We began with convenience sampling, followed by theoretical sampling as the research progressed. Participants were recruited through professional networks \textcolor{black}{such as} LinkedIn and X (formerly Twitter), as well as through personal connections, with a snowballing approach encouraging participants to invite colleagues or friends. We intentionally targeted individuals from diverse backgrounds, including developers at various experience levels and stakeholders in different roles, covering a wide range of geographical locations. The eligibility of each prospective participant was assessed before sharing the pre-interview questionnaire, and those meeting the eligibility criteria received a detailed email outlining the study's requirements. For instance, since the interview focused on interactions between developers and stakeholders, participants who indicated they did not have such interactions were not recruited for the study. 
Data collection involved conducting in-depth, semi-structured interviews with 22 participants (10 developers and 12 stakeholders), using open-ended questions. The interviews were approximately 30-45 minutes long and were conducted via Zoom due to the geographical distribution of participants. The interview guides (see our online Appendix\footnotemark[\value{footnote}]) were employed during the interviews together with appropriate follow-up questions based on the flow of the conversation. Two rounds of data collection took place over five months and all the interviews were audio-recorded. In both rounds, we started interviews with participants defining empathy in general and in the context of working with software developers or stakeholders. Recognising the potential for ambiguity in the interpretation of empathy, this step helped to guide the remainder of the interview and provided a reference point when participants struggled with situational questions. 
\textcolor{black}{In the first round, we began with situational questions, asking about instances where empathy was either present or lacking, then explored the underlying causes and revisited these scenarios to discuss the impact of these experiences.} After analysing the first round data, the interview guides were revised for the second round, prioritising discussions on the impact of empathy or its absence before exploring the causes. The second round also incorporated exploring the strategies identified from the first round and our previous study \cite{gunatilake2024enablers}.

Throughout the interviews, we did not directly ask participants about causes of empathy and the lack thereof, impacts, or strategies. Instead, we prompted them to discuss instances of successful or challenging empathetic experiences. This approach was chosen to allow participants to naturally share insights into empathy without explicitly focusing on these aspects. While this method may have limitations, we believe it contributed to the authenticity of the data we gathered. In addition, we maintained an open line of communication throughout the interviews, encouraging participants to seek clarification on any questions they found unclear or ambiguous.

\subsubsection{Basic Data Analysis} \label{sec:Data Analysis}
\begin{figure}[htbp]
    \centering
    \includegraphics[width=\linewidth]{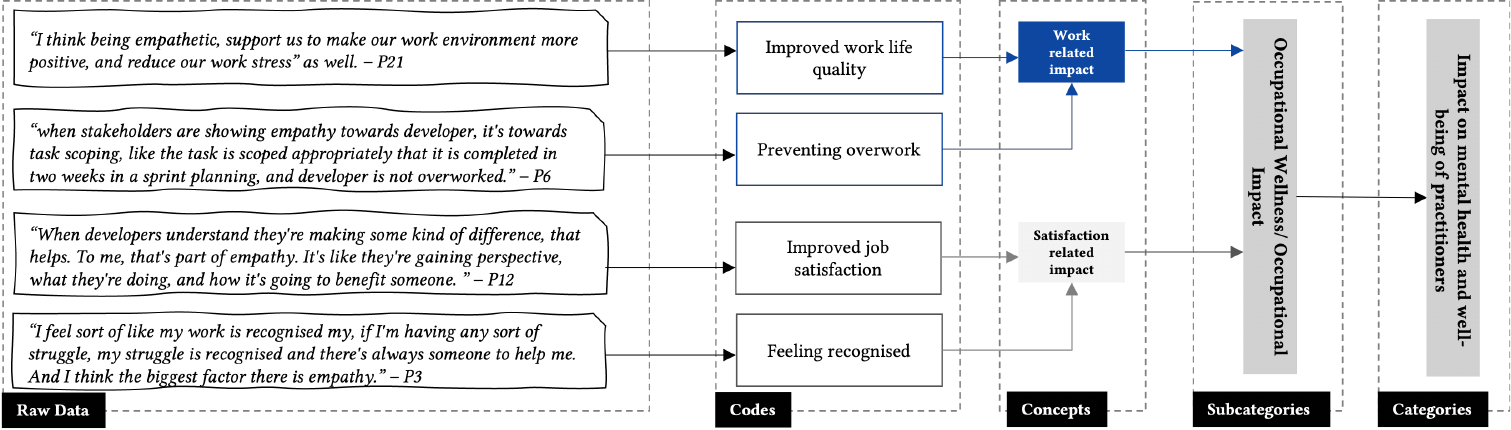}
    \Description{A diagram showing the process of applying STGT}
    \caption{Emergence of the category `\textit{Impact on mental health and well-being of practitioners}' from raw data → codes → concepts → subcategory → category through constant comparison}
    \label{fig:STGT Example}
\end{figure}

After the first round of data collection, we fully analysed all interviews before starting the second round. Throughout the study, data collection was interleaved with ongoing analysis. The interview recordings were transcribed using Otter.ai and stored and analysed using NVivo, following the STGT guidelines \cite{hoda2022STGT, hoda2024qualitative}. The first author conducted open coding on all transcripts, developing a codebook, which was peer-reviewed and refined based on feedback from co-authors, including third author, a grounded theory expert. Similar codes were grouped to form concepts, subcategories, and categories using constant comparison. The first author visualised these groupings, which were further reviewed and refined by the team. An example of our STGT data analysis is illustrated in Figure \ref{fig:STGT Example}. Memos were written to capture researcher reflections, helping document emerging concepts and relationships (see our online Appendix\footnote{https://github.com/Hashini-G/SupplementaryInfoPackage-6CsStudy}). With guidance from the third author, the team gained confidence that key concepts were well-developed enough to proceed to the advanced stage of theory development.

\subsection{Advanced Stage of Theory Development} \label{sec:Theory Development}

In the advanced stage of STGT, researchers can choose between \textit{emergent} and \textit{structured} modes for theory development, both of which can lead to comprehensive theories \cite{hoda2022STGT, hoda2024qualitative}. \textcolor{black}{The emergent mode, based on Glaser's version of GT, allows the theoretical structure to emerge toward the end of the theory development process \cite{hoda2024qualitative}. In contrast, the structured mode, inspired by Strauss and Corbin's approach to GT, involves identifying a structure earlier in the research process to guide subsequent data collection and analysis toward theory development.} The structured mode allows the use of predefined theoretical templates at the beginning of the advanced stage, such as the coding paradigm \cite{strauss1990gt}, conditional matrix \cite{strauss1990gt}, or the Glaserian 6Cs theory template \cite{glaser1978theoretical}, to guide further data collection and analysis. Given that our study mapped well to the 6Cs template, we proceeded with the structured mode of theory development. \textcolor{black}{Glaser's 6Cs theory template is a well-established framework in SE research \cite{graetsch2023data, hoda2011selforganizingagile, madampe2024EI}.} \textcolor{black}{This model organises the central phenomenon or key category, in our case, empathy in developer-stakeholder interactions, with six interconnected categories: Context, Conditions, Causes, Consequences, Contingencies, and Covariances. In this model, the central phenomenon is the focal point of the study. \textit{Context} refers to the setting in which the central phenomenon occurs, while \textit{Conditions} define the factors that shape the central phenomenon. \textit{Causes} are the reasons that lead to the occurrence of the central phenomenon, and \textit{Consequences} describe its effects. \textit{Contingencies} represent the strategies used to address the central phenomenon, and \textit{Covariances} capture the relationships among these other categories. For example, concepts that contributed to empathy or its absence were grouped under \textit{Causes}, while those that described the impact of empathy or lack thereof were categorised as \textit{Consequences}. Similarly, concepts representing strategies to mitigate empathy barriers or enhance empathy were classified under \textit{Contingencies}.} This approach enabled us to systematically align our data collection, analysis, and theory development with the 6Cs template, ensuring that our findings were rigorously integrated within this established structure.


\subsubsection{Structured Data Collection and Analysis}
Using the structured mode, we applied a structured approach to data collection and analysis, aiming to deepen our understanding of the emerging concepts and their relationships to the key category. 
In the second round, we made two key adjustments to our interview approach to align with the structured mode of data collection and analysis. \textcolor{black}{First, we revised the interview guide to focus on adding depth to the identified concepts and categories, particularly the \textit{consequences} category. By prioritising questions on consequences early in the interviews, we were able to allocate more time to this topic, strengthening the existing categories and uncovering new insights. Discussions on \textit{causes} and \textit{contingencies} were moved to later in the interview, as these categories were already well-developed.} Second, through theoretical sampling, we recruited a more diverse group of participants, increasing the number of developers and ensuring cultural and experiential variety. This diversity ensured that our findings captured the nuances of cultural and experiential differences, contributing to a more robust and comprehensive understanding of the studied phenomena.


\subsubsection{Theoretical Saturation and Integration}
According to the STGT guidelines, theoretical saturation is achieved when no new or significantly enhanced concepts, categories, or insights emerge from further data collection \cite{hoda2022STGT, hoda2024qualitative}. After analysing interviews P21 and P22, we concluded that theoretical saturation had been reached, as these interviews only reinforced existing concepts without introducing new findings. 
We applied theoretical integration to ensure these categories were fully integrated into the overall theoretical structure. We identified two key conditions that shape empathy: regular interactions among developers and stakeholders, and other factors that influence empathy. We also found that developer-stakeholder interactions could be further refined by considering the nature of interaction, communication mode, and frequency.
Further, the six concepts under impact on mental health and well-being of practitioners were more effectively grouped into higher-level subcategories: emotional, occupational, and social impact. We also identified opportunities to merge related concepts, such as grouping project planning \& goal setting with impact on agile practices under consequences, and combining customer engagement with stakeholder engagement under contingencies for greater clarity.
This structured approach allowed us to systematically explore and represent the dynamics of empathy within the context of our study.

\subsubsection{Targeted Literature Review (TLR)}
\textcolor{black}{Once we had achieved theoretical saturation and stabilised our concepts and categories}, we conducted a TLR to situate our findings within the broader research landscape. A TLR is an in-depth literature review focused on the relevance of emergent categories and hypotheses, typically performed periodically during the advanced stage of theory development \cite[Chapter 6]{hoda2024qualitative}. This approach enabled us to compare our emergent findings with existing work and identify research gaps. We performed an informal TLR using keywords linked to the categories in our theoretical model, comparing the related works with our findings. The results of this comparison are discussed in Section \ref{sec:Discussion of findings in relation to existing literature}.


\section{\textcolor{black}{Findings - A Theory on the Role of Empathy in Developer-Stakeholder Interactions in SE}} \label{sec:Findings}
\begin{figure}[htbp]
    \centering
    \includegraphics[width=\textwidth]{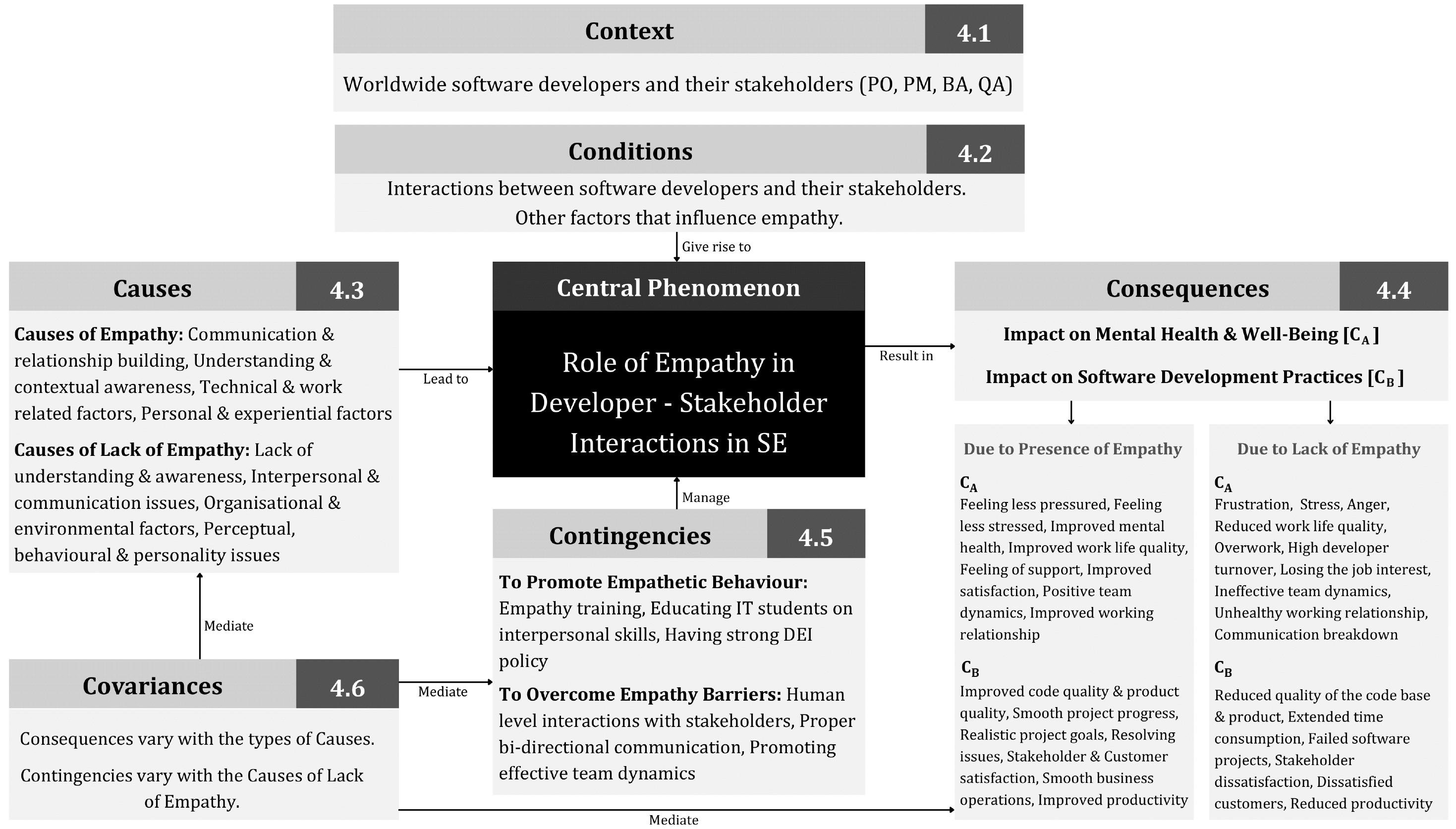}
    \caption{\textcolor{black}{Theory on the role of empathy in developer-stakeholder interactions in SE developed using socio-technical grounded theory (STGT) \cite{hoda2024qualitative} and structured and visualised using the 6Cs template (context, conditions, causes, consequences, contingencies, and covariances) \cite{glaser1978theoretical}. The relevant section numbers are displayed next to the title of each box to indicate the corresponding sections.}}
    \Description{A diagram visualising the theory of the role of empathy in developer-stakeholder interactions in SE structured and visualised using the 6Cs template}
    \label{fig:6Cs}
\end{figure}

We present \textcolor{black}{our theory, structured according to} the 6Cs template \cite{glaser1978theoretical} as illustrated in Figure \ref{fig:6Cs}. Our theory regarding the role of empathy in interactions between developers and stakeholder in SE explains: (a) the \textbf{context}, which outlines key contextual information about our participants, including their work locations, team and organisational contexts of their work; (b) the \textbf{conditions} which explains factors that are prerequisites for the central phenomenon to manifest and \textcolor{black}{alters the central phenomenon in some way}; (c) the \textbf{causes} of presence of empathy and its absence; (d) the \textbf{consequences} of empathy and the lack thereof; (e) the \textbf{contingencies} or strategies applied to enhance empathy or mitigate empathy barriers; (f) the \textbf{covariances}, which represent the relationships between these categories. 
\textcolor{black}{The components of the theory inherently address our RQs, specifically RQ1 is answered through the Context (Section \ref{sec:Context}), Conditions (Section \ref{sec:Conditions}), Causes (Section \ref{sec:Causes}), and Consequences (Section \ref{sec:Consequences}), while RQ2 is addressed in the Contingencies (Section \ref{sec:Contingencies}).}
We share several quotations from participants that serve both as evidence of the underlying raw data and as a way for the readers to experience the phenomenon first hand from the participants' perspective. However, due to confidentiality concerns, we are unable to share all the underlying quotations from our interviews.



\subsection{Context} \label{sec:Context}
\textcolor{black}{In the 6Cs template, context refers to the environment in which the central phenomenon occurs and manifests. In our study, this context specifically pertains to the environment where empathy is expressed and experienced. This context centres on software practitioners, specifically developers and stakeholders, whose professional settings shape how empathy is expressed and experienced through their interactions. Within our theory, context encompasses the demographics, roles, responsibilities, team details, and project settings of developers and stakeholders. These elements collectively define the environment in which empathy unfolds, how it is expressed and perceived in their interactions. To explore this,} we conducted interviews with 22 software practitioners from various regions, with the majority based in Australia (n=8), followed by participants from Asia, the USA, Europe, the UK, and Canada. Table \ref{tab:Demographics of the Participants} provides key demographic information. To maintain participant confidentiality, we refer to them by numbers P1-P10 for developers and P11-P22 for stakeholders throughout this paper. 
The developer group within our participants held various roles, reflecting their diverse levels of experience. Similarly, the stakeholder group represented different stages of the software development life cycle, including roles such as project manager, product owner, testers, and a customer. 
Our participants worked across diverse domains, including finance, healthcare, telecommunication, with some participants working in more than one domain. 
\textcolor{black}{In terms of gender distribution, the sample comprised slightly more female (13/22) than male (9/22) participants.} Participants represented various age groups, with the majority aged 31-40 (n=11) and had varied levels of professional experience, \textcolor{black}{with the majority being experienced practitioners who had more than 5 years of experience.}

\begin{table} [htbp]
    \footnotesize
    \caption{Demographic Information of the Practitioners (P1-P10: Developers, P11-P22: Stakeholders)}
    \label{tab:Demographics of the Participants}
     \setlength{\aboverulesep}{0pt}
    
    \begin{tabular}{P{0.06\linewidth} P{0.2\linewidth} P{0.2\linewidth} P{0.08\linewidth} P{0.07\linewidth} P{0.07\linewidth} P{0.07\linewidth} P{0.07\linewidth}}
        \toprule
         \textbf{ID} & \textbf{Role}  & \textbf{Domain} & \textbf{Country} & \textbf{Age Group} & \textbf{Gender} & \textbf{Experience} & \textbf{\textcolor{black}{Data Col. Round}}\\
         \midrule

         P1 & Software Developer & Biology & UK & 31-40 & Male & 5-10 Y & \textcolor{black}{Basic}\\
         P2 & Software Architect & Telco, CPS & Canada & Above 70 & Male & 20-30 Y & \textcolor{black}{Basic}\\
         P3 & Software Developer & Compiler Engineering & UK & 31-40 & Female & 3-5 Y & \textcolor{black}{Basic}\\
         P4 & Software Developer & Finance & Sweden & 31-40 & Female & 5-10 Y & \textcolor{black}{Basic}\\
         P5 & Software Developer & Automotive & USA & 31-40 & Female & 5-10 Y & \textcolor{black}{Basic}\\
         P6 & Software Developer & Telco, Healthcare, Sales \& Manufacturing, Finance & India & Prefer not to answer & Male & 5-10 Y & \textcolor{black}{Advanced}\\
         P7 & Software Developer & Telco, Healthcare, Finance & India & 20-30 & Male & 5-10 Y & \textcolor{black}{Advanced}\\
         P8 & Lead Technical Consultant & Finance, HRM, TTT & Sri Lanka & 31-40 & Female & 5-10 Y & \textcolor{black}{Advanced}\\
         P9 & Software Developer & TTT, Research & Australia & 20-30 & Female & 3-5 Y & \textcolor{black}{Advanced}\\
         P10 & Senior Software Developer & Insurance, Finance & India & 31-40 & Female & 5-10 Y & \textcolor{black}{Advanced}\\

         
         P11 & Quality Assurance Manager & FSM & USA & 51-60 & Male & 20-30 Y & \textcolor{black}{Basic}\\

         P12 & Product Owner/Manager & FSM, TTT & USA & 31-40 & Male & 5-10 Y & \textcolor{black}{Basic}\\

         P13 & Senior QA Engineer & Finance & Germany & 31-40 & Female & 5-10 Y & \textcolor{black}{Basic}\\

         P14 & Product Owner & Education & Australia & 51-60 & Male & 30-40 Y & \textcolor{black}{Basic}\\

         P15 & Project Manager & Healthcare, Agriculture, Education & Australia & 20-30 & Female & 3-5 Y & \textcolor{black}{Basic}\\

         P16 & Founder/CEO/Customer & Digital Information Storage & Australia & 61-70 & Female & 3-5 Y & \textcolor{black}{Basic}\\

         P17 & Project Manager & Healthcare & Australia & 51-60 & Male & 30-40 Y & \textcolor{black}{Basic}\\

         P18 & Engineering Manager & Telco & Australia & 31-40 & Male & 20-30 Y & \textcolor{black}{Advanced}\\

         P19 & SA/Designer/RE & Charity NPO & Australia & 31-40 & Female & 1-2 Y & \textcolor{black}{Advanced}\\

         P20 & Senior Project Manager & Healthcare, Finance, Law, Manufacturing & Australia & 51-60 & Female & 20-30 Y & \textcolor{black}{Advanced}\\

         P21 & Senior QA Engineer & Restaurant \& Food Delivery & Sri Lanka & 20-30 & Female & 5-10 Y & \textcolor{black}{Advanced}\\

         P22 & Senior QA Engineer & Insurance & USA & 31-40 & Female & 5-10 Y & \textcolor{black}{Advanced}\\

         \bottomrule
    \end{tabular}
    
    \begin{flushleft}   
        \textit{\textcolor{black}{Data Col: Data Collection}, Telco: Telecommunication, CPS: Cyber-Physical Systems, HRM: Human Resource Management, TTT: Transport, Travel \& Tourism, FSM: Field Service Management, SA: Solution Architect, RE: Requirement Engineer, NPO: Non-profit organisation, QA: Quality Assurance. \textcolor{black}{Participants marked as ``Advanced'' were recruited in the second round of data collection using theoretical sampling to increase diversity in cultural background, experience level, and role. ``Basic'' refers to participants from the initial round, recruited through a combination of convenience and theoretical sampling.}}    
    \end{flushleft}
    
\end{table}

\begin{table*}[htbp]
    \caption{Information of Current/Most Recent Project and Team of the Practitioners}
    \label{tab:Information of Current/Most Recent Project and Team of the Participants}
    \resizebox{\textwidth}{!}{
    \begin{tabular}{@{}llllll@{}}
        \toprule
        \textbf{Team Size} & \textbf{\# of Practitioners} & \textbf{Org. Size}& \textbf{\# of Practitioners} & \textbf{Development Method Used} & \textbf{\# of Practitioners}\\
        \midrule
        
        Less than or equal to 5 & \blackbar{4}{4} &  Startup & \blackbar{1}{1} & Agile - Scrum & \blackbar{19}{19}\\
        5 - 10 & \blackbar{11}{11} & Small & \blackbar{2}{2} & Agile - Kanban & \blackbar{7}{7}\\
        10 -20 & \blackbar{3}{3} & Medium & \blackbar{4}{4} & Traditional (Waterfall) & \blackbar{8}{8}\\
        More than 20 & \blackbar{4}{4} & Large & \blackbar{15}{15} & & \\
        
        \cmidrule(r){1-4} \cmidrule(r){5-6}
        \multicolumn{2}{@{}l}{\textbf{Job Responsibilities}} & \multicolumn{2}{@{}l}{\textbf{Frequency*}} & \textbf{Self Affinity (Human vs Technology)} & \textbf{\# of Practitioners}\\
        \cmidrule(r){1-4} \cmidrule(r){5-6}

        \multicolumn{2}{@{}l}{Req. gathering \& elicitation with proxy users} & \multicolumn{2}{@{}l}{\multicolorbar{6}{4}{7}{2}{3}{[6, 4, 7, 2, 3]}} & Predominantly Human-Centric  & \blackbar{7}{7}\\
        
        \multicolumn{2}{@{}l}{Req. gathering \& elicitation with real end-users} & \multicolumn{2}{@{}l}{\multicolorbar{4}{3}{5}{6}{4}{[4, 3, 5, 6, 4]}}  & Somewhat Human-Centric & \blackbar{9}{9}\\
        
        \multicolumn{2}{@{}l}{Designing software (UI/UX)} & \multicolumn{2}{@{}l}{\multicolorbar{1}{4}{8}{4}{5}{[1, 4, 8, 4, 5]}} & Predominantly Technology-Centric & \blackbar{3}{3}\\
        
        \multicolumn{2}{@{}l}{Front-end development/Programming} & \multicolumn{2}{@{}l}{\multicolorbar{4}{2}{4}{2}{10}{[4, 2, 4, 2, 10]}} & Somewhat Technology-Centric & \blackbar{3}{3}\\

        \cmidrule(r){5-6}
        \multicolumn{2}{@{}l}{Back-end development/Programming} & \multicolumn{2}{@{}l}{\multicolorbar{7}{2}{3}{2}{8}{[7, 2, 3, 2, 8]}} & \textbf{Team's Affinity (Human vs Technology)} & \textbf{\# of Practitioners}\\
        \cmidrule(r){5-6}

        \multicolumn{2}{@{}l}{Testing} & \multicolumn{2}{@{}l}{\multicolorbar{8}{8}{5}{1}{0}{[8, 8, 5, 1, 0]}} & Predominantly Human-Centric  & \blackbar{1}{1}\\
        
        \multicolumn{2}{@{}l}{Fixing Defects} & \multicolumn{2}{@{}l}{\multicolorbar{5}{4}{4}{2}{7}{[5, 4, 4, 2, 7]}} & Somewhat Human-Centric & \blackbar{8}{8}\\
        
        \multicolumn{2}{@{}l}{User Testing} & \multicolumn{2}{@{}l}{\multicolorbar{4}{5}{7}{3}{3}{[4, 5, 7, 3, 3]}} & Predominantly Technology-Centric & \blackbar{7}{7}\\
        
        \multicolumn{2}{@{}l}{Usability Testing} & \multicolumn{2}{@{}l}{\multicolorbar{5}{1}{8}{6}{2}{[5, 1, 8, 6, 2]}} & Somewhat Technology-Centric & \blackbar{6}{6}\\

        \multicolumn{2}{@{}l}{Conducting User Training} & \multicolumn{2}{@{}l}{\multicolorbar{3}{4}{3}{3}{9}{[3, 4, 3, 3, 9]}} & & \\
        \multicolumn{2}{@{}l}{User Support Services} & \multicolumn{2}{@{}l}{\multicolorbar{3}{3}{6}{2}{8}{[3, 3, 6, 2, 8]}} & & \\
        \multicolumn{2}{@{}l}{Technical Writing/User Documentation} & \multicolumn{2}{@{}l}{\multicolorbar{2}{6}{7}{1}{6}{[2, 6, 7, 1, 6]}} & & \\
        \multicolumn{2}{@{}l}{Maintenance and Operations} & \multicolumn{2}{@{}l}{\multicolorbar{7}{4}{4}{4}{3}{[7, 4, 4, 4, 3]}} & & \\
            
        \bottomrule
    \end{tabular}
    }

    \begin{flushleft}
         \textit{*Order of the bars in Frequency column of Job Responsibilities graph: Always → Very Often → Sometimes → Rarely → Never, followed by respective values.}
    \end{flushleft}

\end{table*}

All the participants were working in team contexts and shared their experiences by referring to their current or most recent projects. The relevant team and project information are summarised in Table \ref{tab:Information of Current/Most Recent Project and Team of the Participants}. 
Most of their teams were more technology-centric, whereas they themselves tended to have a stronger affinity for human-centric approaches. 
Some practitioners had experience in both developer and stakeholder roles. In our pre-interview questionnaire, we asked participants to self-identify the role that best described them. Those with dual-role experience chose which role to focus on during the interviews. 
We encountered only one instance where a participant with experience in both roles sought guidance on role selection. In this case, we considered two factors: the years of experience in each role and the recentness of their experience. Since their experience was balanced between both roles, we asked them to focus on their more recent stakeholder role, as those experiences were still fresh in the memory.

\subsection{Conditions} \label{sec:Conditions}
\textcolor{black}{Conditions describe the factors that shape the central phenomenon. Our findings indicate that the empathy among practitioners is influenced by two main conditions: the \textit{interactions between developers and stakeholders}, and \textit{other factors that influence empathy} by shaping how it is perceived and expressed.}
%
\textbf{Regular interactions between developers and stakeholders} was a condition for empathy to emerge \textcolor{black}{(P1-P22)}. 
Without meaningful and sustained interactions, the opportunity for empathy to manifest was significantly diminished. Therefore, the regular interactions themselves were a foundational condition necessary for empathy to emerge. We identified that the \textit{nature of interactions}, \textit{mode of communication}, and \textit{frequency of interactions} helped to better describe these regular interactions. The nature, mode, and frequency of these interactions created the environment in which empathy can either thrive or falter. 

The \textit{nature of interactions} refers to the specific instances where practitioners engage. Participants described it by referring to different meetings or various stakeholder interactions \textcolor{black}{(P1-P22)}. These meetings included requirements discussions, feature/defect discussions, design reviews, and technical meetings. Participants highlighted collaborative activities during project management, and agile ceremonies including sprint planning, daily stand-up, and retrospective. Other interactions involved problem identification, troubleshooting, documentation, and working with other teams. These interactions occurred with various stakeholders. 
Such collaborative engagements, often centred around shared goals, were crucial for fostering empathy.
%
The \textit{mode of communication} refers to the media or various methods used for interaction \textcolor{black}{(P1,P4-P22)}. Participants discussed face-to-face, virtual, and asynchronous communication modes. Most commonly, they mentioned virtual meetings facilitated through platforms \textcolor{black}{such as} Zoom, Google Meet, and MS Teams. Participants also highlighted in-person meetings, emails, LinkedIn messaging, and communication via ticketing systems such as JIRA. 
\begin{quote}
    \small
    \faIcon{comments} \textit{``Mode of communication is basically anything you could possibly think of. So obviously these days a lot via Teams internally, typing in Teams, as well as having both audio and video calls with the Teams, as well as in person when I'm actually in the office. Occasionally even old-fashioned telephone and SMS, if the case arises.'' 
    - P18 \textcolor{black}{(Developer)}}
\end{quote}

The \textit{frequency of interaction} refers to how often developers and stakeholders engaged \textcolor{black}{(P1, P2, P4-P22)}, ranging from daily stand-ups and email exchanges to weekly meetings with product owners and monthly retrospectives. Participants also highlighted ad-hoc meetings and regular interactions, such as sprint planning and reviews, and sessions with UI/UX designers. These consistent engagements built rapport and deeper understanding, fostering empathy between practitioners.


\textcolor{black}{We identified the \textbf{factors that influence empathy} as the other condition. Participants highlighted various factors that influence how empathy is perceived and expressed, including personality (P1, P2, P11, P15, P16, P17, P20), culture (P4, P13, P15, P19, P22), job role (P2, P14, P17), and gender (P15).
\textcolor{black}{Participants emphasised that \textit{personality} significantly influences empathy in SE (P1, P2, P11, P15, P16, P17, P20). Several participants highlighted personality as a factor shaping the capacity for empathy. For instance, P16 suggested that ``you’re either an empathetic person or not,'' linking empathy to stable personality traits and arguing that it extended beyond mere interactional skill. This view was echoed by others who believed that empathy often stemmed from an individual’s nature (P1, P20), though it also required interpersonal trust and mutual respect to be sustained (P16).
Participants also discussed how personality influenced responses to emotional expression in workplace settings. For example, P11 reflected on instances where team members became emotionally vulnerable, only to later regret having opened up, leading them to withdraw or even respond defensively in future interactions. These emotional dynamics, shaped by individual disposition and workplace norms, could challenge the maintenance of empathetic relationships.
Several participants perceived that developers who identify as more introverted (P1, P2, P16) may sometimes prioritise solitary work over interpersonal engagement, which can make empathetic exchanges more difficult in collaborative settings. Separately, other participants noted that a strong focus on technical problem-solving or task completion (P2, P15, P17) may lead some developers to de-emphasise relational aspects of teamwork, even if unintentionally. However, these views were not universally held, and participants acknowledged that empathy manifests differently depending on individual dispositions and situational factors.}
\textbf{Cultural context} also influenced empathy (P4, P13, P15, P19, P22). Participants who worked in different countries observed variations in how empathy was expressed and encouraged. 
Open and inclusive company cultures fostered empathy, while rigid and hierarchical structures hindered it.
Participants noted that empathy varied between different \textbf{job roles} (P2, P14, P17). Tensions often arose between developers and project managers due to conflicting priorities, and testers faced challenges in maintaining empathy with developers during defect identification. 
\textbf{Gender diversity} was another factor (P15). Teams with gender diversity, especially including female members, tended to have more open communication and regular meetings, which facilitated empathy. In contrast, all-male teams, particularly those in back-end roles, were seen as less communicative and less likely to foster empathetic interactions.}
\begin{quote}
    \small
    \faIcon{comments} \textit{``in terms of people who are most asocial of all the engineers, technical people, software people take the cake, because they're so focused on what they do. And the discipline attracts this kind of people. These are people who are introverts, which doesn't mean that they can't empathise. But if they're introverts, that's almost a necessary condition for lack of empathy. Not true as a general statement, but it's one of the conditions. So they don't want to talk to customers. Literally, I've seen people say, I don't want to talk to them, or these people are not smart enough to understand what I'm doing, therefore I have no respect for them, and so on. So I do want to emphasise, at least that's my experience from all this time is that, software in particular, in terms of empathy, it's got a huge problem. It's a problem, and therefore it has to be studied, and properly understood. And then maybe some mitigations can be devised that deals with this problem of introverts, asocial software developers.'' - P2 (Developer)}
\end{quote}

\begin{quote}
    \small
    \faIcon{comments}\textit{\textcolor{black}{``I think probably the difficulty that they have is that they get so focused in their own work. So for example, I'm working on some other projects work outside of the project that the developers are involved on. And I try to, what I would really like is for them to participate in some kind of thought input into the work that I'm doing. But often, I don't get that. Because they are so narrowly focused on what they're doing, that they can't look outside their own area. So I think that's a bit of a barrier there is that people get too insular in their own work, and therefore, they don't really care about what's going on outside their own area.'' - P17 (Stakeholder)}}
\end{quote}

\subsection{Causes} \label{sec:Causes}
\textcolor{black}{In this study, we explored both the causes of presence of empathy and its absence. Participants were asked to share their experiences of both showing empathy and struggling to be empathetic during their interactions. From these scenarios, we inquired about the factors that enabled or hindered their ability to be empathetic, and we identified several causes that contributed to fostering empathy and those that impeded it.} 

\begin{figure}[htbp]
    \centering
    \includegraphics[scale=0.35]{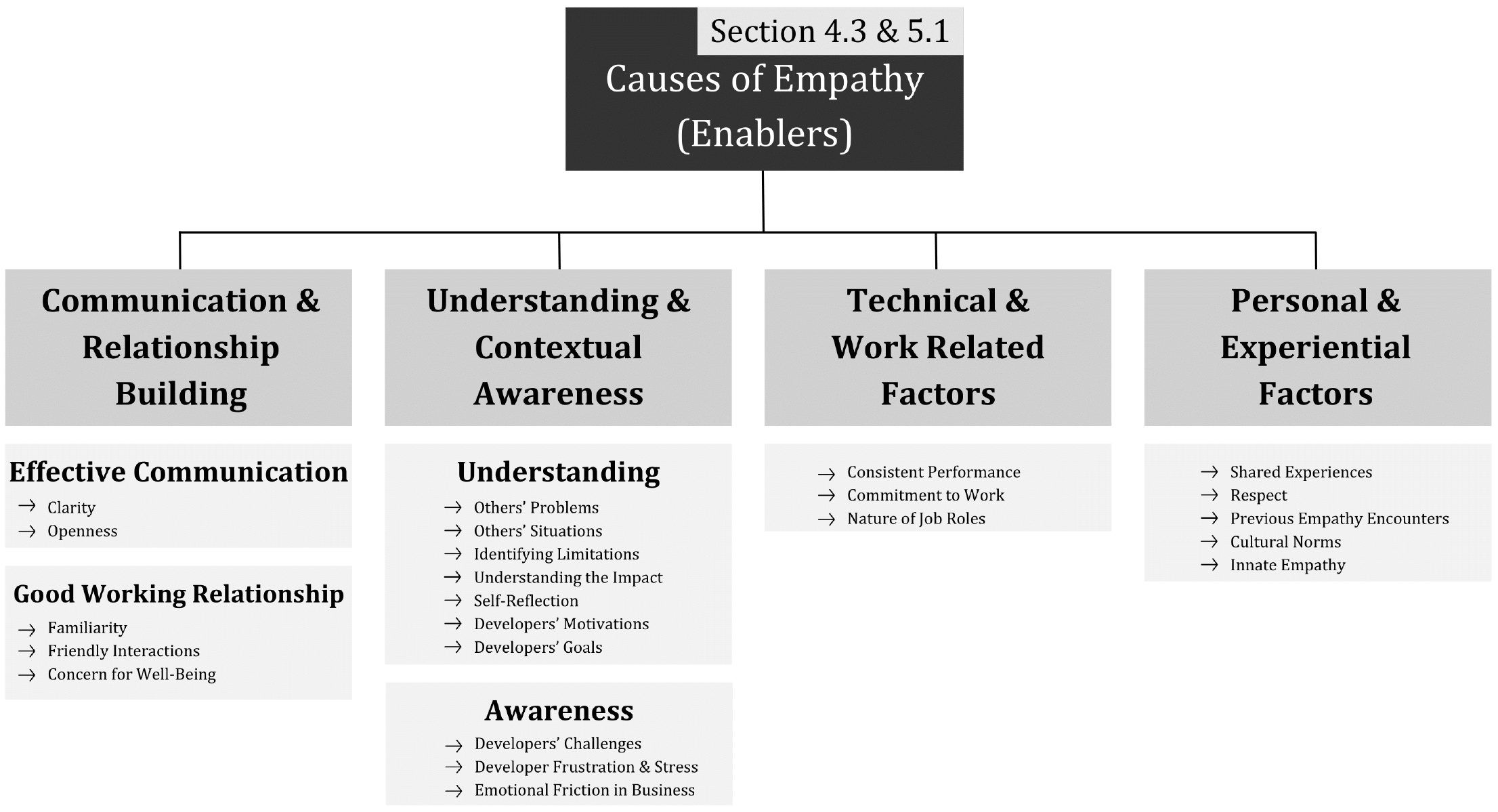}
    \Description{A diagram showing the categorisation of the causes of empathy}
    \caption{Causes of Empathy}
    \label{fig:Causes of Empathy}
\end{figure}

\textcolor{black}{We identified six concepts under the \textit{causes of empathy}: effective communication, good working relationship, understanding, awareness, technical and work related factors, and personal and experiential factors, as illustrated in Figure \ref{fig:Causes of Empathy}. 
Each concept reflects distinct contributors to empathy: \textit{effective communication} highlights the role of clear and open dialogue; \textit{good working relationships} emphasise strong interpersonal connections; understanding involves deep comprehension of others’ problems and limitations; \textit{awareness} relates to recognising emotional and practical challenges faced by developers; \textit{technical and work-related factors} concern professional expertise and performance; and \textit{personal and experiential factors} draw on individual experiences, cultural norms, and shared goals. These concepts are discussed in detail in Section \ref{sec:Causes of Empathy (Enablers)}.}

\begin{figure}[htbp]
    \centering
    \includegraphics[scale=0.35]{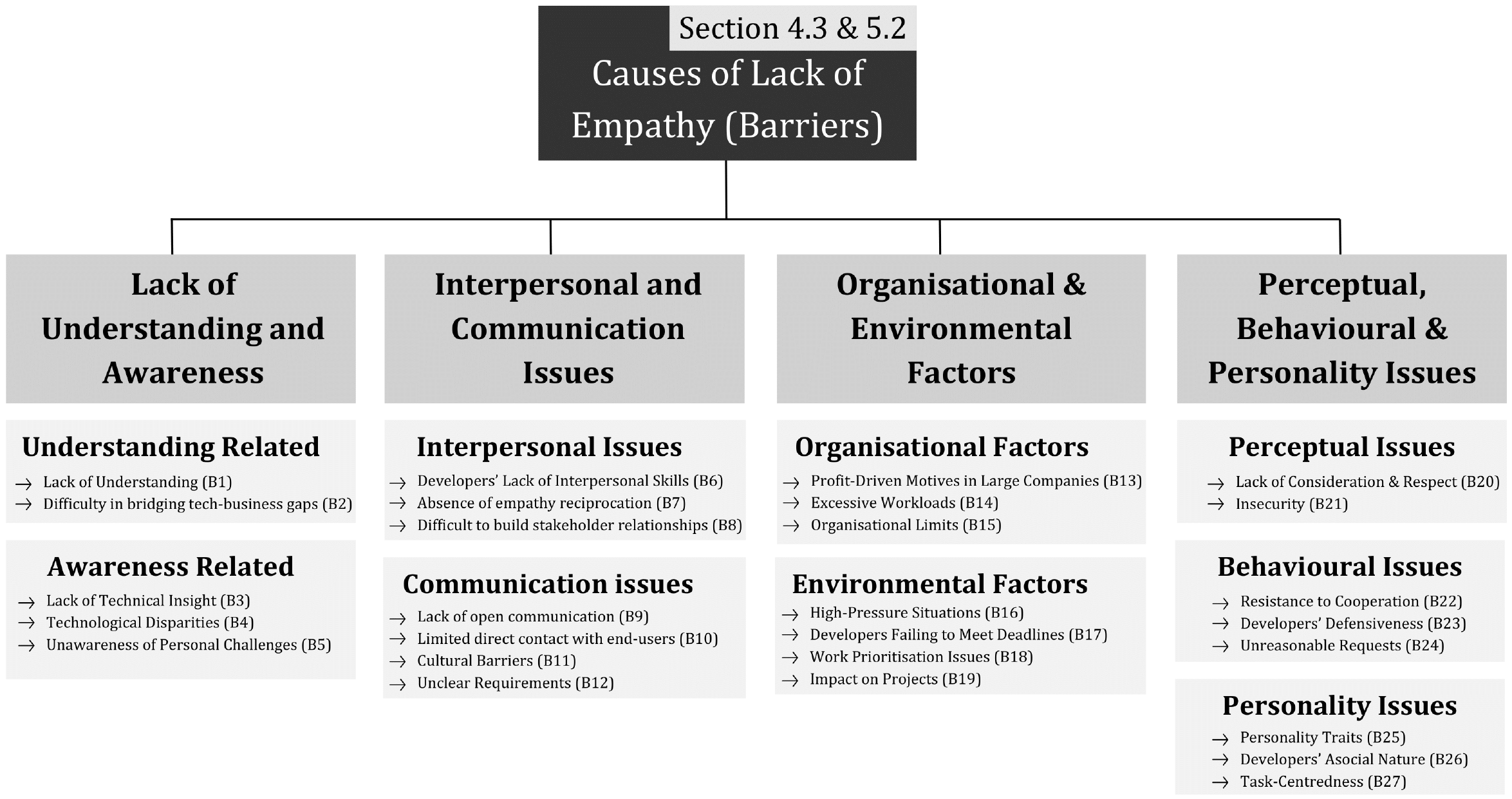}
    \Description{A diagram showing the categorisation of the causes of lack of empathy}
    \caption{Causes of Lack of Empathy}
    \label{fig:Causes of Lack of Empathy}
\end{figure}

\textcolor{black}{We identified nine key concepts under the \textit{causes of lack of empathy}: challenges in understanding, lack of awareness, interpersonal issues, communication related issues, organisational factors, environmental factors, perceptual issues, behavioural issues, and personality issues, as illustrated in Figure \ref{fig:Causes of Lack of Empathy}. 
\textit{Challenges in understanding} encompasses barriers that arise due to lack of knowledge or understanding of the other party's work, technical constraints, and experiences. \textit{Lack of awareness} captures instances where empathy was hindered by insufficient awareness of others’ technical constraints, and experiences. \textit{Interpersonal issues} relate to difficulties arising from strained professional interactions, while \textit{communication-related issues} stem from ineffective or limited communication. Organisational factors highlight broader contextual influences such as structural constraints or business priorities. \textit{Environmental factors} refer to workplace conditions that affect empathetic engagement. \textit{Perceptual issues} involve individual interpretations that undermine empathy, \textit{behavioural issue}s reflect uncooperative or defensive conduct, and \textit{personality issues} relate to traits that inhibit empathetic responses. These concepts are discussed in detail in Section \ref{sec:Causes of Lack of Empathy (Barriers)}.}

\subsection{Consequences of Empathy (Impact of Empathy)} \label{sec:Consequences}
\textcolor{black}{Consequences describe the positive and negative impacts of empathy in SE. We identified both expected and ``unexpected'' consequences of empathy on developer-stakeholder interactions. By ``unexpected'' we refer to findings that were surprising or did not naturally align with the typical notion of empathy as a prosocial behaviour, where its presence led to negative outcomes and its absence resulted in positive ones. These unexpected findings are discussed in detail in Section \ref{sec:Negative Impact of Presence of Empathy} and Section \ref{sec:Positive Impact of Lack of Empathy}. 
Conversely, ``expected findings'' are those that aligned with the conventional view of empathy as a prosocial behaviour, where its presence led to positive outcomes and its absence to negative ones. Within these expected consequences, we identified two main categories: the impact on mental health and well-being of practitioners, and the impact on software development practices.}

\begin{figure}[htbp]
    \centering
    \includegraphics[scale=0.35]{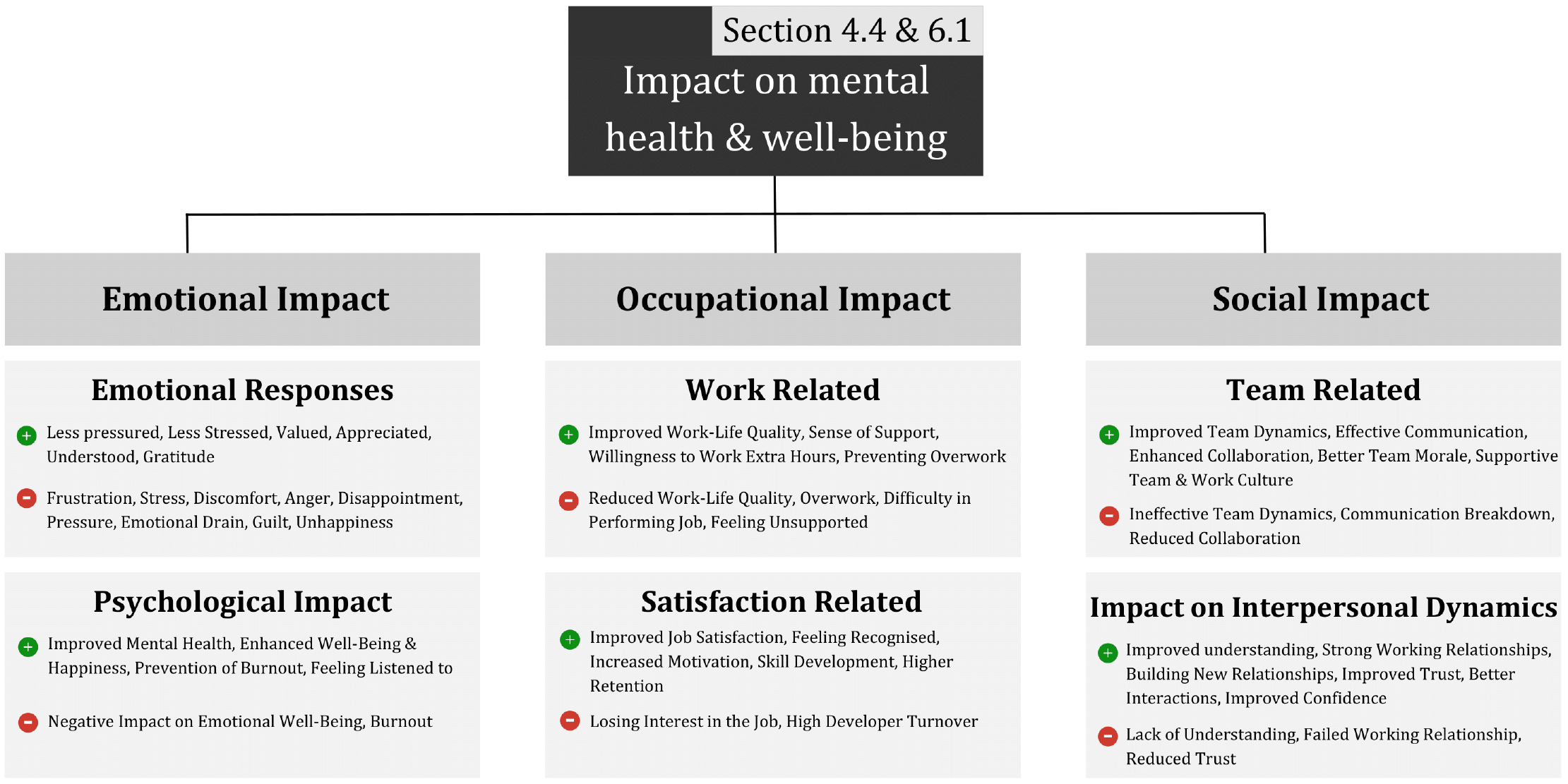}
    \Description{A diagram showing the categorisation of the impact on mental health and well-being of practitioners due to empathy}
    \caption{Positive \& negative impact on mental health \& well-being of practitioners due to presence \& absence of empathy}
    \label{fig:Impact on mental health}
\end{figure}

\textcolor{black}{\textit{Impact on mental health and well-being of practitioners} category encompasses the various ways in which the work environment, tasks, and organisational culture in software development affect the mental health and well-being of practitioners. We identified both positive impacts stemmed from the presence of empathy, and negative impacts resulted from its absence. These consequences aligned with three of the six categories in Hettler's wellness model: emotional, occupational, and social wellness \cite{hettler1984wellness}. In this paper, we refer to these categories as \textit{emotional impact, occupational impact,} and \textit{social impact}, as illustrated in Figure \ref{fig:Impact on mental health}. 
\textit{Emotional impact} refers to the immediate emotional responses and broader psychological effects practitioners experience in their work environment due to the presence or absence of empathy. Participants  described a range of \textit{emotional responses} including various emotional reactions such as happiness, frustration, and anxiety, as well as longer-term \textit{psychological impacts} such as burnout, depression, as well as enhanced psychological resilience and overall well-being.
The \textit{occupational impact} focuses on how empathy affects job-related factors such as work-life balance, job security, and professional growth. Participants described two main areas: \textit{work-related impact}, reflecting how empathy affects occupational health; and \textit{satisfaction-related impact}, referring to the extent to which empathy contributes to a sense of fulfilment and job satisfaction.
\textit{Social impact} captures the influence of empathy on workplace relationships and social interactions. 
Participants discussed \textit{team related impact}, focusing on how empathy or its absence in team dynamics influences practitioners' social wellness. This includes the quality of interactions within the team, team cohesion, and support systems among team members. They also discussed \textit{impact on interpersonal dynamics} examining how empathy affects the quality and nature of interpersonal relationships at work. It includes communication effectiveness, collaboration, and the sense of community and belonging among colleagues.
These findings are discussed in detail in Section \ref{sec:Impact on mental health and well-being of practitioners}.}

\begin{figure}[htbp]
    \centering
    \includegraphics[scale=0.35]{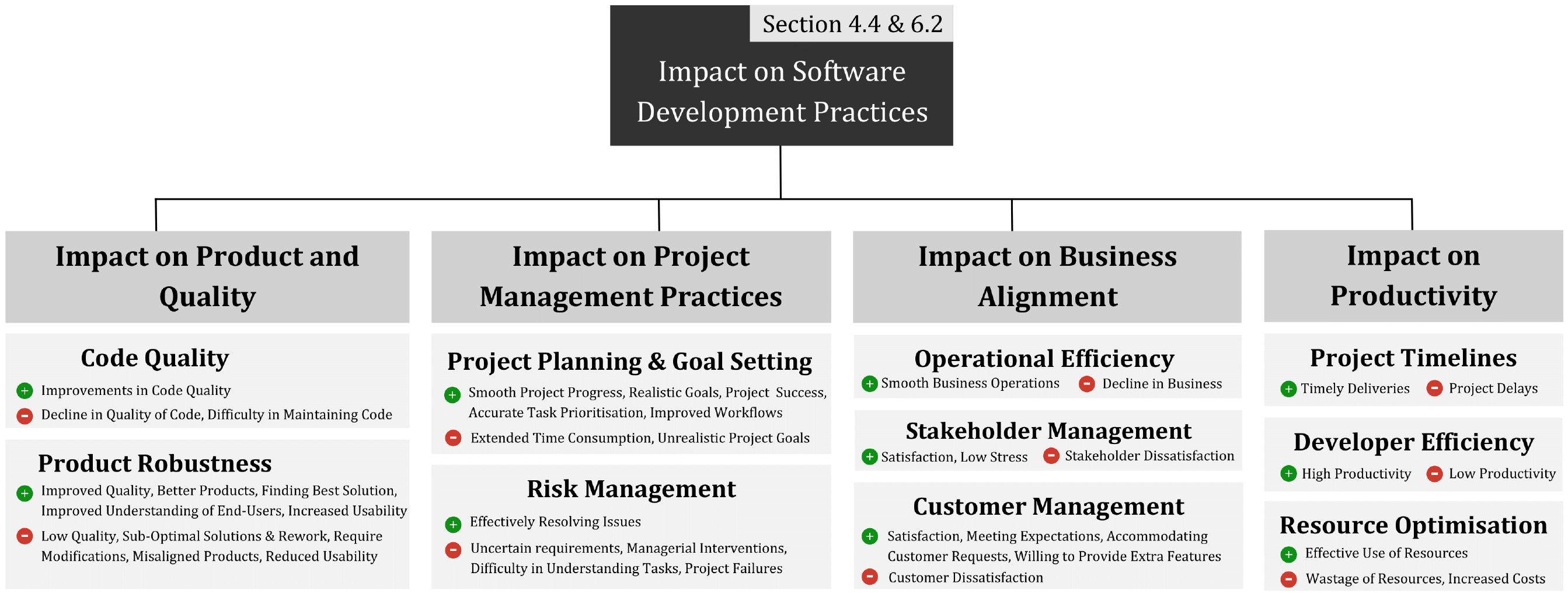}
    \Description{A diagram showing the categorisation of the impact on software development practices due to empathy}
    \caption{Positive \& negative impact on software development practices due to presence \& absence of empathy}
    \label{fig:Impact on software development practices}
\end{figure}

\textcolor{black}{\textit{Impact on software development practices} examines how empathy affects various aspects of software development practices that influence the overall success of software projects. We identified the positive and negative impacts stemmed from the presence of empathy and its absence, respectively. These consequences are categorised as \textit{impact on product and quality, impact on project management practices, impact on business alignment} and \textit{impact on productivity}, as illustrated in Figure \ref{fig:Impact on software development practices}. 
\textit{Impact on product and quality} focuses on how empathy affects the technical quality of the software. Participants discussed impact on code quality and product robustness under this category. The \textit{code quality} addresses the technical quality of the code, including aspects such as maintainability, readability, reliability, and absence of defects. The \textit{product robustness} refers to the resilience and stability of the product in various conditions, including its performance, scalability, and ability to handle errors and unexpected inputs.
\textit{Impact on project management practices} focuses on how empathy influences project outcomes. Participants discussed impact on both project planning \& goal setting, and risk management under this category. The \textit{project planning \& goal setting} addresses practices related to project planning, goals, and agile practices ensuring that they are aligned with both team capabilities and business objectives. The \textit{risk management} involves identifying, assessing, and mitigating risks to ensure successful project completion.
\textit{Impact on business alignment} addresses how empathy helps align development efforts with broader business objectives. Participants discussed impact on operational efficiency, stakeholder management, and customer management under this category. The \textit{operational efficiency} focuses on improving overall processes and workflows within the organisation or project to streamline operations, and minimise inefficiencies. The \textit{stakeholder management} aims at effectively understanding, engaging with, and satisfying the needs of stakeholders. The \textit{customer management} involves the strategic management of relationships with customers or end-users. 
\textit{Impact on productivity} highlights how empathy affects practitioners’ efficiency and productivity. Participants discussed impact on project timelines, developer efficiency, and resource optimisation under this category. The \textit{project timelines} addresses the ability to deliver products within the agreed timelines, ensuring that deadlines are met. The \textit{developer efficiency} refers to how effectively developers can perform their tasks, including the use of tools and practices that enhance productivity. The \textit{resource optimisation} involves the efficient use of available resources, such as time, skills, and budget, to maximise output and minimise waste.
These findings are discussed in detail in Section \ref{sec:Impact on Software Development practices}.}

\subsection{Contingencies (Strategies)} \label{sec:Contingencies}

\begin{figure}[htbp]
    \centering
    \includegraphics[scale=0.35]{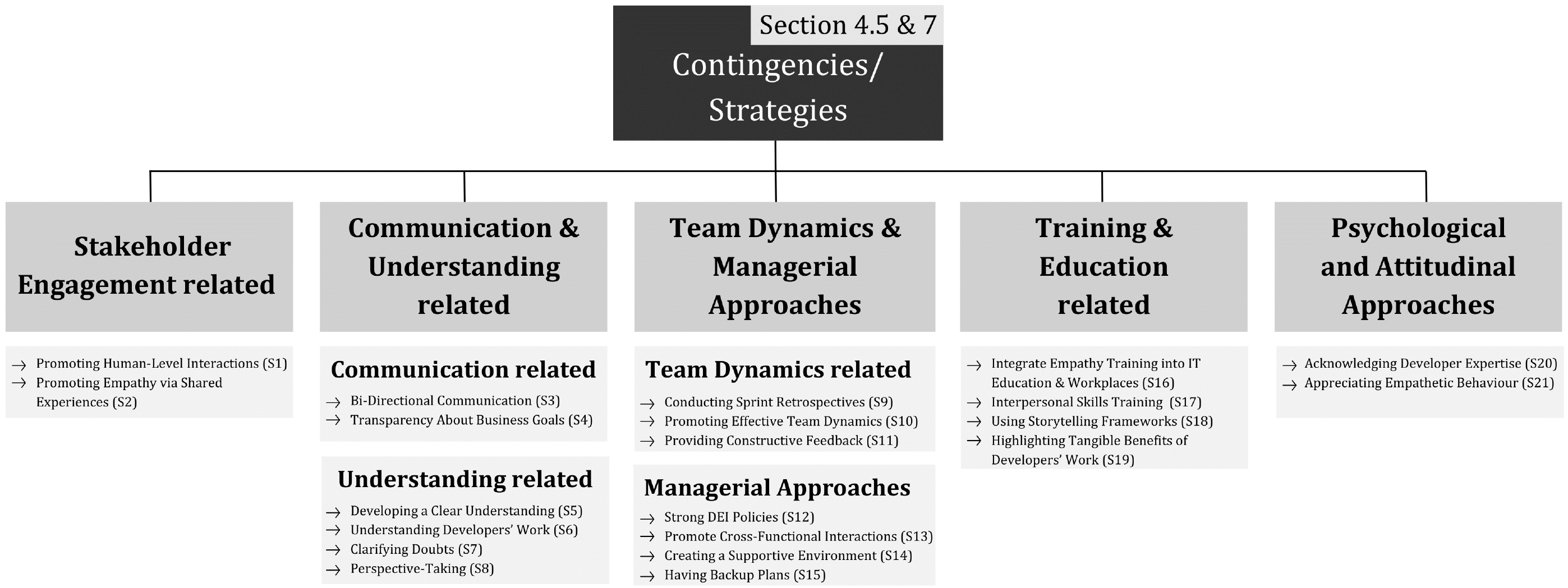}
    \Description{A diagram showing the types of empathy related strategies}
    \caption{Empathy related strategies}
    \label{fig:Strategies}
\end{figure}

\textcolor{black}{Contingencies encompass various strategies that software practitioners employ to foster, maintain, or navigate empathy in their professional interactions. While the term `Contingencies' is used in the original 6Cs template, we adopt the term `Strategies' in this article, as it more accurately reflects the nature of the actions described. We identified five concepts under the strategies of empathy including the strategies related to customer and stakeholder engagement, communication and understanding, team dynamics and managerial approaches, training and education, and psychological and attitudinal approaches, as illustrated in Figure \ref{fig:Strategies}.} 

\textcolor{black}{The \textit{customer and stakeholder engagement} concept includes strategies aimed at improving interactions and relationships with stakeholders, ensuring their perspectives were understood and effectively incorporated into project development. \textit{Communication and understanding} involves strategies for enhancing the quality and effectiveness of communication among practitioners, fostering a deeper mutual understanding of needs, perspectives, and goals. \textit{Team dynamics and managerial approaches} encompass strategies aimed at fostering empathy through team dynamics and management practices. \textit{Training and education} includes strategies that focus on training and educating practitioners to enhance their skills and knowledge, particularly in areas that improve empathy. Finally, \textit{psychological and attitudinal approaches} encompass strategies that encourage mindset shifts and attitudinal changes to create a more empathetic and psychologically safe team environment. These findings are discussed in detail in Section \ref{sec:Elaboration of the Contingencies}.}

\subsection{Covariances} \label{sec:Covariances}
\textit{Covariance} refers to the relationship between two categories, where changes in one category correspond to changes in the other \cite{glaser1978theoretical, hoda2011selforganizingagile}. 
We identified two key instances of covariance: (A) consequences (one category) varied depending on the types of causes (the other category), and (B) contingencies varied based on the causes of lack of empathy. 

\textbf{Covariance A.} \textit{Consequences varied depending on the types of causes.} We identified two main types of causes: causes of presence of empathy and causes of its absence. The consequences associated with each type of causes varied accordingly. We found both positive and negative consequences in relation to the presence and absence of empathy. Section \ref{sec:Impact on mental health and well-being of practitioners} and Section \ref{sec:Impact on Software Development practices} outline these consequences, organised into two key categories: mental health and well-being of practitioners, and software development practices. In addition, Section \ref{sec:Negative Impact of Presence of Empathy} details the negative consequences of presence of empathy, while Section \ref{sec:Positive Impact of Lack of Empathy} highlights the positive consequences of lack of empathy. 
 
\textbf{Covariance B.} \textit{Contingencies varied based on the causes of lack of empathy.} We identified specific contingencies aimed at mitigating the causes of empathy deficits. \textcolor{black}{The mapping between the strategies and causes of lack of empathy was established through a combination of participant input and data analysis. Some strategies were directly articulated by the participants during the interviews, while for others, we applied STGT to systematically analyse and identify relationships between the strategies and causes. This approach ensured that the mappings were grounded in the data and reflected both direct participant feedback and inferred relationships.} For each of these causes, we identified contingencies that could be applied to overcome these challenges. For instance, we identified that one cause of lack of empathy, \textit{difficulty in bridging technical-business gap} could be mitigated by several contingencies, including having proper bi-directional communication (S3), clarifying doubts (S7), and promoting professional \& cross-functional interactions (S13). We summarised the empathy contingencies in Table \ref{tab:Empathy Contingencies} and these covariances in Table \ref{tab:Covariances Between Causes of Lack of Empathy and Contingencies}. 

\begin{table*}[htbp]
    \centering
    \footnotesize
    \caption{Covariances Between Causes of Lack of Empathy and Contingencies \textcolor{black}{(Strategies)}}
    \label{tab:Covariances Between Causes of Lack of Empathy and Contingencies}
     \begin{tabular}{P{0.03\textwidth} P{0.4\textwidth} P{0.25\textwidth}}
        \toprule
        \textbf{ID} & \textbf{Causes of Lack of Empathy} & \textbf{Contingencies \textcolor{black}{(Strategies)}} \\
        \midrule
        B1 & Lack of understanding & S1, S3, S4-S7, S14, S18, S19\\
        B2 & Difficulty in bridging technical-business gaps  & S3, S7, S13 \\
        B3 & Lack of technical insight & S3, S6-S8, S13\\
        B4 & Technological disparities & S7, S13\\
        B5 & Unawareness of personal challenges & S3, S10, S14\\
        B6 & Developers' lack of interpersonal skills & S3\\
        B7 & Absence of empathy reciprocation & S8, S10, S21 \\
        B8 & Difficulty in building positive stakeholder relationships & S1-S3, S5, S8-S10\\
        B9 & Lack of open communication & S3, S4\\
        B10 & Limited direct contact with end-users & S1, S3, S7, S18, S19\\
        B11 & Cultural barriers & S3\\
        B12 & Unclear requirements & S3, S4, S13\\
        B13 & Profit-driven motives in large corporations & S3, S4, S6, S10\\
        B14 & Excessive workloads & S3, S4, S6, S7, S10, S20\\
        B15 & Organisational limits & S3-S6, S10\\
        B16 & High-pressure situations & S1, S3-S6, S8, S10 \\
        B17 & Developers failing to meet deadlines & S1, S3-S6, S8, S10, S15\\
        B18 & Work prioritisation issues & S3, S4, S5-S8, S14\\
        B19 & Impact on projects & S1, S3, S10, S14, S15 \\
        B20 & Lack of consideration and respect & S1, S3, S5-S8, S10, S20\\
        B21 & Feelings of insecurity & S1-S5, S9-S11, S14, S21 \\
        B22 & Resistance to cooperation & S1, S3, S7, S9-S11, S13-S15, S21\\
        B23 & Developers' defensiveness & S1, S3, S5, S6, S10, S11, S13, S14, S20\\
        B24 & Unreasonable requests & S3, S4, S5, S6, S7, S9, S20\\
        B25 & Personality traits & S2, S3, S9, S10, S20, S21\\
        B26 & Developers' asocial nature & S2, S3, S8, S9, S10, S21\\
        B27 & Task-centredness & S1-S3, S7, S9, S10\\
        \bottomrule
     \end{tabular}
\end{table*}

\section{\textcolor{black}{Elaboration of the Causes of Empathy and Its Absence}} \label{sec:Elaboration of the Causes of Empathy and Its Absence}
\textcolor{black}{As introduced in the theory overview (Section \ref{sec:Findings}), we identified several causes for the presence and absence of empathy. In this section, we elaborate on them in detail.}

\subsection{Causes of Empathy (Enablers)} \label{sec:Causes of Empathy (Enablers)}
\textbf{Empathy due to effective communication} emerged as a key factor in fostering empathy during professional interactions. Participants emphasised the importance of \textit{clarity of communication} (P1, P3, P8, P9) and value of \textit{openness when interacting with others} (P7-P9, P11, P15, P17-P19, P21).
Developers highlighted the significance of \textit{clear communication} in fostering empathy. They shared that simplifying technical and business language helps to bridge the understanding gaps, which promoted empathy between developers and stakeholders. 

\begin{quote}
    \small
    \faIcon{comments} \textit{``I was asking some detailed questions about some bugs that leaked into production. I could feel the developer getting nervous or a little bit defensive. I said, listen I'm not doing this to place blame. I just want to understand from a technical perspective, so I can see how things should be in the future, how can we help each other in the future. As soon as I put that out there, he immediately calmed down. And he immediately said, Okay, I get it,  I'm sorry, I understand where you're coming from. And we had a great conversation over slack. And he helped me, and I helped him.'' - P11 \textcolor{black}{(Stakeholder)}}

\end{quote}

\textbf{Empathy due to good working relationships} was supported by participants’ emphasis on the role of strong interpersonal connections in fostering empathy. Contributing factors included \textit{familiarity} (P2, P3, P5, P7, P9-P14, P18), \textit{friendly interactions} (P2, P13, P16, P17, P21), and \textit{concern for well-being} (P4, P5, P17). 
\textit{Familiarity} with team members was a significant contributor to empathy, as knowing each other's work patterns, communication styles, and personal backgrounds helped individuals understand each other's pressures and challenges more deeply.
\textit{Maintaining a professional and friendly relationship} also fostered empathy, with participants noting that positive working dynamics made individuals more likely to go an extra mile for each other. 
\begin{quote}
    \small
    \faIcon{comments} \textit{``So overall, I think that probably at the end of the day, you know, actually having a courteous, friendly and frequent interaction. So it wasn't just sort of, you know, okay, here's your stuff that we've delivered, I've delivered for you this week [..] It was a discussion, it was a warm discussion, there was warmth in the relationship.'' - P16 \textcolor{black}{(Stakeholder)}}
\end{quote}

\textbf{Empathy due to understanding} was reflected in participants’ emphasis on the importance of developing a deep understanding of others. This included recognising others' problems (P2, P5, P7-P10) and situations (P8, P10, P20, P22), identifying limitations (P13, P14, P18), understanding the impact (P6, P10, P12), self-reflection (P9, P19), grasping developers' motivations (P12, P22), and understanding developer goals (P13, P19). 
Many participants shared how empathy emerged when they took time to understand others' \textit{problems} and \textit{situations} by putting themselves in others' shoes. 
Developers' awareness of \textit{potential real-world impact of their work} also promoted empathy. 
\textit{Self-reflection} emerged as a key factor in developing empathy. Participants noted that reflecting on situations, both during (reflection-in-action) and after interactions (reflection-on-action), allowed them to better understand others' experiences and respond more empathetically \cite{schon2017reflective}. 
\begin{quote}
    \small
    \faIcon{comments} \textit{``You can create empathy by showing someone who's very value and problem oriented, how much time you can save someone by doing X thing. So that's usually what I'll try to do. We can usually just say like, if we do this, these people will save 30 minutes a day. 30 minutes a day is a big deal.'' - P12 \textcolor{black}{(Stakeholder)}}
\end{quote}

\textbf{Empathy due to awareness} was evident in participants’ reflections on how recognising emotional and practical challenges contributed to empathetic responses. This included being aware of developers' challenges (P6, P15, P18, P21), recognising developer frustration and stress (P18, P19, P21), and understanding the emotional friction in business (P2, P11). 
Participants observed that empathy was fostered by \textit{recognising the challenges developers faced}, especially those beyond their core expertise. 
Participants noted that empathy helped to reduce inherent \textit{emotional friction in business settings}. They observed that by recognising the negative emotions that could build up and approaching situations empathetically, they were able to break down these barriers. This awareness of emotions and their impact on interactions aligns closely with the principles of emotional intelligence \cite{goleman2002emotional}, suggesting that empathy in SE may be underpinned by emotional awareness and regulation.
\begin{quote}
    \small
    \faIcon{comments} \textit{``I show empathy to them \textcolor{black}{[developers]} to break down friction, that wall, which is inherently there in the nature of our business. My philosophy is there's no anger in business. It's only business, there should be no emotion in business. When people starting to get emotional, usually it's anger or frustration, or sadness in some cases, then I feel we need to defuse that and get into empathetic mode, and get on the same page, and we can have a conversation.'' - P11 \textcolor{black}{(Stakeholder)}}
\end{quote}

\textbf{Empathy due to technical and work related factors} was reflected in participants’ recognition of how professional competence and work behaviours influenced empathetic responses. Contributing factors included consistent performance (P5, P10, P16), commitment to work (P1, P13, P16), and nature of job roles (P1, P9).
\textit{Consistent and dedicated performance} from developers was highlighted as a key factor in fostering empathy. 
Their strong track record and willingness to go an extra mile led stakeholders to be more empathetic towards them. Empathy was linked to the \textit{commitment shown by team members}. When stakeholders or developers noticed others' genuine dedication to delivering quality results, it promoted a shared sense of empathy. 
\begin{quote}
    \small
    \faIcon{comments} \textit{``We worked on it and he worked in it, obviously things went wrong. So it's easy to have more empathy for a person who's trying to do their best and trying to understand your needs, and willing to fix things when they need to be fixed in the build, than if someone goes, oh there it is, if you want more you have to pay or something like that.'' - P16 \textcolor{black}{(Stakeholder)}}
\end{quote}

\textbf{Empathy due to personal and experiential factors} was shaped by participants’ reflections on how individual backgrounds and experiences fostered empathy. This included shared experiences (P2, P3, P7, P9, P14, P20), respect (P2, P19, P20), previous experience and encounters with empathy (P4, P10, P13), cultural norms (P3, P4), and innate empathy (P2, P7, P16). 
\textit{Shared experiences} including both job-related and personal, fostered empathy. 
This sense of shared experience helped to bridge the gaps in understanding and facilitated more empathetic interactions. 
Participants highlighted that \textit{cultural norms} and team culture played a significant role in shaping empathetic interactions. In some cultures, there was a strong emphasis on supporting each other which naturally encouraged empathy during personal emergencies. Some described \textit{empathy as an intrinsic quality} in certain individuals, noting that empathy can stem from a natural disposition, although it can also be cultivated.
\begin{quote}
    \small
    \faIcon{comments} \textit{``[..] one of our support engineers brought up a question, it was an offshoot of a customer's question, and what he wanted to do was sort of like grow his knowledge within our product base, and to grow his skills. And, for me, what I felt was like, I like to ask a lot of questions to, again to grow my knowledge, even if it doesn't seem relevant to what I'm doing right now. So I was able to empathise with him there. And because of that, rather than thinking of his question is not relevant, his question is not important, because it's his question and not a customer question, I was able to reach out to the other engineers in my team as well and sort of answer his question and help him explore and broaden his knowledge a bit more.'' - P3 \textcolor{black}{(Developer)}}

\end{quote}

\subsection{Causes of Lack of Empathy (Barriers)}  \label{sec:Causes of Lack of Empathy (Barriers)}
\textbf{Challenges in understanding} emerged as a barrier to empathy when participants described difficulties in grasping others’ work, experiences, or technical constraints. These included a general lack of understanding (P3, P6, P9, P14, P15) and difficulty in bridging technical-business gaps (P1, P12, P15).
Many participants noted that a fundamental \textit{lack of understanding} was a cause of lack of empathy. Stakeholders unfamiliar with the complexities of software development often had unrealistic expectations, leading developers to feel overlooked. Stakeholders noted that developers were often disconnected from end-users, which limited their understanding of end-users. 
\begin{quote}
    \small
    \faIcon{comments} \textit{``because they don't understand what exactly we do for work. For example, they wouldn't know how to code or to produce a piece of software or anything. So I guess in their heads, it's more like, oh, they're able to do anything, they are able to code anything. So I guess, that kind of takes away empathy, takes away looking at us like it's a human being behind this piece of software.'' - P3 \textcolor{black}{(Developer)}}
\end{quote}

\textbf{Lack of awareness} was described by participants as a barrier to empathy stemming from limited insight into others’ experiences or constraints. This included a lack of technical insight (P1, P2, P9, P18), unawareness of personal challenges (P13, P22), and technological disparities (P3, P7).
A lack of empathy also arose from the \textit{unawareness of technological disparities}. Developers shared how they initially struggled to empathise with stakeholders who lacked technological access or skills, as the developers themselves were proficient in modern technologies. 
\begin{quote}
    \small
    \faIcon{comments} \textit{``my lack of empathy came with the fact that I was very technologically proficient in the sense that, I had the skills to use a smartphone or to use any sort of equipment. But most of the staff who were under the customer hadn't given access to a smartphone [..] it was hard for me to wrap my head around the fact that, this is 2019 2018, and there are people who still don't know how to use a smartphone.'' - P3 \textcolor{black}{(Developer)}}
\end{quote}

\textbf{Interpersonal issues}  were noted by participants as barriers to empathy arising from challenges in professional interactions. These included the absence of empathy reciprocation (P2, P7, P10, P11, P16), developers' lack of interpersonal skills (P12, P16), and difficulty in building positive stakeholder relationship (P2, P5).
Empathy was often described as a reciprocal process, with participants noting that the \textit{absence of reciprocation} often led to a breakdown in empathetic interactions. When empathy was not returned, individuals were less likely to continue showing it. 
\begin{quote}
    \small
    \faIcon{comments} \textit{``when the team has communicated to stakeholders that this is not the way it happens, constantly they have been pushing us, not being empathetic [..] there's always a breaking point for everyone. So after a point you realise that, I have no business of being empathetic to them. Because even after open communication and several iterations of doing so, if it is not reciprocated, then we take that stand that I'm not going to be empathetic any more.'' - P7 \textcolor{black}{(Developer)}}
\end{quote}

\textbf{Communication related issues} were identified as significant barriers to empathy. Participants highlighted factors such as lack of open communication (P6, P9, P13, P15), limited direct contact with end-users (P14, P18), cultural barriers (P4, P13), and unclear requirements (P1, P6). 
The \textit{absence of open communication} impeded the development of empathy. Participants described instances where individuals could not express their concerns, resulting in misunderstandings and making it difficult for others to show empathy.

\begin{quote}
    \small
    \faIcon{comments} \textit{``When they \textcolor{black}{[developers]} have no empathy with me, it's no communication. They are kind of not very interested in talking in the meeting [..] they just want to finish that task. They don't want to listen to other things whether I am on track or not. They don't care about products.'' - P15 \textcolor{black}{(Stakeholder)}}
\end{quote}

\textbf{Organisational factors} were identified as contributors to a lack of empathy within the broader organisational context. These included organisational constraints (P8, P20, P22), profit-driven motives in large corporations (P2, P7), and excessive workloads (P7, P10).
\textit{Organisational boundaries}, such as contractual obligations and time constraints, limited the ability to show empathy. Participants described situations where they had to refuse providing support due to strict adherence to budgets, timelines, or predefined roles. 
\begin{quote}
    \small
    \faIcon{comments} \textit{``So maybe they had their own pressures, and have committed things, maybe their situation has been, I can't afford to be empathetic at this situation, I want to get things done [..] as a core, people are empathetic. I do believe in it. But I think the situation that they are in, maybe they don't express it, they're not in a situation to do that [..]'' - P7 \textcolor{black}{(Developer)}}
\end{quote}

\textbf{Environmental factors} were identified as barriers related to the working environment. These included high-pressure situations (P5, P7, P9, P13), work prioritisation issues (P9, P14), developers failing to meet deadlines (P16, P20), and impact on projects (P7, P10). 
\textit{High-pressure scenarios}, such as tight deadlines, made it difficult for participants to consider others' perspectives. Due to stress, they focused solely on task completion, with empathy taking a backseat. 
When the \textit{success of a project was at stake}, empathy diminished. Developers recounted instances where the project's critical importance or resource constraints led stakeholders to prioritise outcomes over empathy.
\begin{quote}
    \small
    \faIcon{comments} \textit{``So the first situation would be the criticality of the project, and then the other main problem was I was the only resource who understand that system, and that would have an impact on other things as well. So the main focus is on, like it might affect the productivity. So that's one barrier which affected empathy.'' - P10 \textcolor{black}{(Developer)}}
\end{quote}

\textbf{Perceptual issues} were identified as barriers rooted in individual perceptions, including lack of consideration \& respect (P2, P3), and insecurity (P21, P22). 
The \textit{lack of consideration and respect} hindered empathy. Some stakeholders viewed developers as capable of coding anything, disregarding the required effort. This led developers to question stakeholders' understanding of software complexities hindering empathetic interactions.
\begin{quote}
    \small
    \faIcon{comments} \textit{``it essentially has a lot has to do with respect. If your customer is being pigheaded and is asking for things, which are may be unreasonable or either too early or they're asking for something which is much more work than they think, then they are not being reasonable. So just I think that, the sense that your opinion is not being respected, that creates a distance and empathetic divergence.'' - P2 \textcolor{black}{(Developer)}}
\end{quote}

\textbf{Behavioural issues} were identified as barriers stemming from individual behaviours, including unreasonable requests (P2, P7, P8, P10), developers' defensiveness (P18, P22), and resistance to cooperation (P11, P12). 
%
When developers viewed feedback as criticism, they became \textit{defensive}, hindering perspective-taking and disrupting empathy. In addition, stakeholders mentioned instances where developers were \textit{resistant to change}, which limited empathetic interactions. 
\begin{quote}
    \small
    \faIcon{comments} \textit{``I had a hard time being empathetic with that team. they immediately told me what we're doing doesn't work for them [..] they came to the meeting with an agenda to shut me down and shut everybody else who was arguing for common processes [..] it is very difficult to have empathy for someone who doesn't want to hear, doesn't want to cooperate, and has already decided that whatever you have to say is worthless before you even open your mouth.'' - P11 \textcolor{black}{(Stakeholder)}}
\end{quote}

\textbf{Personality issues} were identified as barriers related to individual personality traits, including task-centredness (P2, P15, P17), personality traits (P12, P22), and developers' asocial nature (P2, P19).
Developers' \textit{intense focus on their specific tasks} was noted as a barrier to empathy. This narrow view led them to disregard the needs and perspectives of others outside their immediate scope of work. 
Many developers, particularly those who were \textcolor{black}{\textit{with introverted or asocial personality traits}}, preferred working independently and avoiding external communication, which led to a lack of engagement with stakeholders and hindered empathy. 
\begin{quote}
    \small
    \faIcon{comments} \textit{``that's just their personality. They were prototypical, stereotypical, really good developer like they got to be where they were at. As in a senior individual contributor position, because they were really good. They were great problem solvers, very efficient, opinionated on the technical aspect of their work. And the thing with that is, naturally just like that rewards that behaviour [..] and I think that's what's worked for that developer and they have no interest in like doing the business side or managing people.'' - P12 \textcolor{black}{(Stakeholder)}}
\end{quote}

\section{\textcolor{black}{Elaboration of the Consequences of Empathy and Its Absence}} \label{sec:Elaboration of the Consequences of Empathy and Its Absence}
\textcolor{black}{As introduced in the theory overview (Section \ref{sec:Findings}), we identified several consequences of  empathy and its absence. In this section, we elaborate on them in detail.}

\subsection{Impact on Mental Health and Well-being (Expected Consequences of Empathy)} \label{sec:Impact on mental health and well-being of practitioners}
%
Participants discussed \textbf{emotional responses} highlighting \textit{positive emotional responses} such as feeling less pressured (P3, P8, P21, P22), less stressed (P9, P20, P21), valued (P16, P18, P21), appreciated (P14, P16), understood (P20, P21), gratitude (P9, P10), as well as 
\textit{negative emotional responses} such as frustration (P1-P3, P5, P7, P9, P12, P14, P16, P18, P20-P22), stress (P6-P8, P13, P21, P22), discomfort (P3, P5, P16, P17, P20), anger (P11, P14, P16, P18), disappointment (P8, P10, P16, P17), pressure (P3, P13, P21), emotional drain (P12, P18), guilt (P10, P22), and unhappiness (P8, P10).
Participants noted that empathy from team members helped to \textit{alleviate pressure}. 
Some developers even \textit{expressed gratitude} towards stakeholders who showed empathy during critical situations. 
Participants frequently experienced \textit{frustration} due to a lack of empathy. Developers were frustrated when their work was devalued, or their input was not considered in decision-making processes. 
\begin{quote}
    \small
    \faIcon{comments} \textit{``I was thankful to him. I understand it's a hard deadline, that's like a million dollar project, we can't easily compromise with clients when we delay the product. So it's a major decision. it's not easily they allow another developer to pitch in and do the work. So I was really grateful and thankful to him when he understood the situation.'' - P10 \textcolor{black}{(Developer)}}

\end{quote}

Participants discussed \textbf{psychological impacts} by emphasising the \textit{positive psychological impact} of empathy, including improved mental health (P5, P7, P18, P22), enhanced well-being \& happiness (P11, P12), prevention of burnout (P3, P6), feeling listened to (P9, P21), and \textit{negative psychological impact} including negative impact on emotional well-being (P5, P7, P11) and burnout (P6, P7).  
Practitioners shared that empathy led to \textit{improved mental health} by reducing feelings of isolation, making them feel more connected and supported during stressful periods. 
Developers reported a \textit{decline in emotional well-being} when empathy was absent, leading to reduced productivity and difficulty focusing on coding tasks.
\begin{quote}
    \small

    \faIcon{comments} \textit{``first thing was, this is not the best place to be. There's already a lot of myths and jargons about software corporate that, it's a place which is very toxic compared to other industries, there's cut-throat competition, there's always work, no work life balance [..] So first feeling was, they don't understand me, and I think if you're not valued and understood here, then what is even the point in doing all this, let me just do something else and look out somewhere else.'' - P7 \textcolor{black}{(Developer)}}
\end{quote}

Participants discussed \textbf{work related impact} including \textit{positive work related impact} such as improved work-life quality (P1, P7, P11-P13, P15, P18, P19, P21, P22), a sense of support (P3, P4, P7-P9, P15, P18, P20-P22), team members' willingness to work extra hours to meet deadlines (P10, P11), and prevention of overwork (P5, P6, P21). 
The \textit{negative work related impact} included reduced work-life quality (P1, P4, P7, P8, P13, P14, P18), overwork or additional workload (P6, P21), difficulty in performing job tasks (P8, P9), and feeling unsupported (P8, P10).
Empathy contributed to an \textit{improved work-life quality} by fostering clear expectations and mutual understanding, making tasks easier to manage. 
Absence of empathy led to a \textit{deterioration in work-life quality}, making work harder to manage and causing frustration. This impacted their ability to deliver quality results and made everyday tasks feel more tedious.
\begin{quote}
    \small

    \faIcon{comments} \textit{``he [developer] was very less empathetic or not at all empathetic with me in that situation. So it did affect my work life balance. I was always stressed that I need to get this done at this event. otherwise he would report me or he would talk rude to me, scold me like that. I was losing my quality of life. Even my sleep was disturbed and the next day I go with this baggage, like, yesterday this happened. So I was not focused on the next day also.'' - P13 \textcolor{black}{(Stakeholder)}}
    
\end{quote}

Participants discussed \textbf{impact on satisfaction} including \textit{positive impact on satisfaction} such as improved job satisfaction (P1, P10, P12, P15), feeling recognised (P3, P5, P8), increased motivation (P10, P17, P22), enhanced skill development (P6, P19), and higher retention (P3, P12). 
The \textit{negative impact on satisfaction} included losing interest in the job (P1, P2, P6, P7, P22), and high developer turnover (P6, P7, P10, P18).
\textcolor{black}{Participants noted that \textit{increased job satisfaction} was a key outcome of empathy in the workplace, as it enhanced interpersonal relationships and task fulfilment. Empathy also boosted \textit{motivation}, leading to greater enthusiasm and improved work quality. Conversely, when stakeholders fostered an unsupportive work environment, \textit{developers were often motivated to leave}. Developers specifically mentioned declining future opportunities with companies that had previously failed to demonstrate empathy.} 
\begin{quote}
    \small
    \faIcon{comments} \textit{``when we show empathy towards people, they get motivated. for me it's the nature. When someone is doing something good for me, I want to do better for them. That's why I think most of the people behave in that manner.'' - P22 \textcolor{black}{(Stakeholder)}}
   

    
\end{quote}

Participants discussed \textbf{team related impact} including \textit{positive team related impact} such as improved team dynamics (P2, P4, P5, P7-P13, P15, P18-P22), effective communication (P1, P3, P4, P7, P8, P12, P14, P19, P21), enhanced collaboration (P1, P4, P9, P12, P16, P19, P20), better team morale (P9, P17, P18), supportive team and work culture (P3, P13, P17, P22). 
The \textit{negative team related impact} included ineffective team dynamics (P1, P7-P10, P11, P13, P15, P18, P19, P21, P22), communication breakdown (P2, P4, P15, P16, P19, P22), and reduced collaboration (P15, P17).
Empathy improved \textit{team dynamics} by promoting mutual understanding and trust. 
Ability to rely on teammates during high-pressure situations, created a supportive environment, resulting in more cohesive teams.
The lack of empathy \textit{disrupted team dynamics}, causing misunderstandings, fragmented relationships, and inefficient, isolated work.
\begin{quote}
    \small
    \faIcon{comments} \textit{``When team members are empathetic towards each other and other teams, morale is high, people enjoy working together, [..] experience of developing a product has a better feel, people enjoy working on the product. When people are not empathetic, when they're antagonistic, and when they're difficult, it's a very negative experience.'' - P11 \textcolor{black}{(Stakeholder)}}

\end{quote}

Participants discussed \textbf{impact on interpersonal dynamics} including \textit{positive impact on interpersonal dynamics} such as improved understanding (P2, P4, P7-P11, P17, P18, P21), strong working relationships (P4, P7, P9, P11, P13, P16, P19, P21), building new relationships (P2, P5, P7, P19), improved trust (P4, P7, P12), better interactions (P1, P2), and improved confidence (P13, P15).  
The \textit{negative impact on interpersonal dynamics} included lack of understanding (P4, P9), failed working relationship (P1, P7, P8, P10, P13, P16), and reduced trust (P17, P21).
Participants frequently noted that empathy \textit{improved understanding}. It helped team members grasp perspectives and challenges faced by their colleagues, strengthening relationships and collaboration.
Empathy deficits damaged professional relationships, reducing trust and collaboration, with one stakeholder sharing that it led to the termination of a working relationship.

\begin{quote}
    \small
    \faIcon{comments} \textit{``first one was a sense of belonging that people trust me here, people understand me. And it is not just stakeholder developer relationship, that signs of more like a comfort. So you feel that, okay this is something where I belong, people trust me, and I would want to also repay this type of behaviour, maybe when I'm put in that situation.'' - P7 \textcolor{black}{(Developer)}}

\end{quote}

\subsection{\textcolor{black}{Impact on Software Development practices (Expected Consequences of Empathy)}} \label{sec:Impact on Software Development practices}

Participants discussed the impact on \textbf{code quality} including \textit{positive impact on code quality} such as improvements in code quality (P3, P22), and \textit{negative impact on code quality} such as a decline in overall quality of the code base (P1, P6, P7) and increased difficulty in maintaining it (P1, P6). 
Participants described how empathy \textit{helped to improve code quality}. Empathy fostered a collaborative approach to coding and testing, with developers incorporating additional safeguards to address issues identified during empathetic interactions with customers. These interactions helped developers better understand the difficulties customers faced, motivating them to enhance the code. 
Developers noted that when stakeholders lacked empathy, they struggled with unclear requirements and lacked a full understanding of the broader purpose of their tasks. This led to working without a strong sense of direction, resulting in a \textit{decline in the overall quality of the code}. 
\begin{quote}
    \small
    \faIcon{comments} \textit{``earlier how we \textcolor{black}{[QAs]} worked was, once the developers finish coding, they'll send that over to us. They don't understand what QA does, or they didn't mind knowing that. But with that scenario, later on when it comes to finding bugs before sending it to QA, they were able to fix it, they were able to identify it earlier than in the QA process.'' - P22 \textcolor{black}{(Stakeholder)}}

\end{quote}

Participants discussed impact on \textbf{product robustness} including \textit{positive impact on product robustness} such as improved product quality (P2, P3, P6-P8, P10-P14, P16-P22), development of better products (P1, P3, P5-P7), helping to find the best technical solution (P4, P8, P9, P11), improved understanding of end-users (P9, P12, P14, P18), and increased usability (P13, P18). 
The \textit{negative impact on product robustness} included reduced product quality (P2, P4-P7, P12-P14, P21, P22), sub-optimal solutions \& rework (P6, P12, P15), need for modifications to developed features (P1, P9), products that failed to meet user needs (P14, P18), and reduced usability (P1, P3).
Empathy fostered strong collaboration between developers and stakeholders, resulting in the \textit{development of high-quality products}. Empathy encouraged developers to focus on delivering better results considering user perspectives to identify and address issues early. 
The lack of empathy caused stress, frustration, and disengagement among developers, which led to rushed, poorly executed solutions, ultimately \textit{decreasing product quality} leading to bugs.
\begin{quote}
    \small
    \faIcon{comments} \textit{``it was very buggy product and customer didn't like it, and they reached out to sending email like, this was really buggy and they didn't like it, so it was not good experience.'' - P22 \textcolor{black}{(Stakeholder)}}

\end{quote}

Participants discussed the impact on \textbf{project planning \& goal setting} including \textit{positive impact on project planning \& goal setting} such as smoother project progress (P1, P6, P8, P12, P15, P21), more realistic goals (P3, P4, P9), achieving successful project outcomes (P13, P21), accurate task prioritisation (P14, P15), and improving workflows (P15, P19). 
The \textit{negative impact on project planning \& goal setting} included extended time consumption (P9, P10) and unrealistic project goals (P1, P9, P10). 
Empathy facilitated \textit{smoother project progress} by aligning developers and stakeholders with project goals and deadlines. Both groups recognised that understanding each other's challenges increased the likelihood of meeting timelines. 
\textit{Unrealistic goals} were imposed without understanding the team's workload, leading to rushed work, burnout, and decreased quality. 
\begin{quote}
    \small
    \faIcon{comments} \textit{``deadlines are all about understanding the capability of people in the team, limitations, situation people are in, and their pressures. if you're not aware of that, then it can definitely be difficult to settle on a deadline that's successful for everyone, with lack of empathy, you're more likely to set up deadlines that won't be successful.'' - P9 \textcolor{black}{(Developer)}}

\end{quote}

Participants discussed the impact on \textbf{risk management} including \textit{positive impact on risk management} such as effectively resolving issues (P8, P10, P17, P19), and \textit{negative impact on risk management} including uncertainty of requirements (P1, P6), managerial interventions (P11, P17), difficulty in understanding development tasks (P1, P18), and sometimes failed software projects (P16). 
Empathy played a crucial role in the \textit{effective resolution of critical issues} by fostering open communication and mutual understanding. In one case, a project manager empathised with an overwhelmed developer, stepping into resolve a mistake that could have affected the entire project, ensuring smooth delivery without disruptions. 
In another case, the lack of empathy and understanding between developers and stakeholders resulted in a \textit{project failure}, with the entire project collapsing due to strained relationships.
\begin{quote}
    \small
    \faIcon{comments} \textit{``there was a situation where I had to deal with multiple work at the same time. I had to switch between different work. While working on multiple work, I actually made a mistake in one of the work. That had an impact on everyone working in it. My project manager understood my situation and he helped me to get out of that situation.'' - P8 \textcolor{black}{(Developer)}}

\end{quote}

Participants discussed the impact on \textbf{operational efficiency} including \textit{positive impact on operational efficiency} such as smooth business operations (P1, P10), and \textit{negative impact on operational efficiency} such as potential decline of the business (P7, P10).
Developers shared that empathy helped to ensure the end product met client expectations, \textit{fostering seamless business operations}. Empathy-driven collaboration helped businesses avoid delays, meet deadlines, and maintain client satisfaction. 
Participants shared that absence of empathy led to missed deadlines and a breakdown of customer trust, which could result in the \textit{decline of the business}. 
Another developer noted that lack of empathy from stakeholders resulted in missed opportunities for potential revenue.
\begin{quote}
    \small
    \faIcon{comments} \textit{``I also have shown my performance in terms of monetary [..] I was developing a tool where one client was using it, and then after the tool was getting good, there were 20 more clients who on-boarded to use that same tool. So maybe another developer can come and work on it, but still, maybe I would have done it better, since I know it from earlier, that the time could be saved on this transferring things, KT [knowledge transfer] and everything.'' - P10 \textcolor{black}{(Developer)}}
\end{quote}

Participants discussed the impact on \textbf{stakeholder management} including \textit{positive impact on stakeholder management} such as increased stakeholder satisfaction (P1, P5-P7, P19, P20), reduced stress (P3, P18), and \textit{negative impact on stakeholder management} such as stakeholder dissatisfaction (P1, P7, P9, P10, P14). 
Empathy encouraged workload sharing among team members, \textit{reducing pressure on stakeholders} and improving team morale. 
The absence of empathy resulted in \textit{dissatisfaction} due to poor communication and a lack of understanding of stakeholders' needs. 
\begin{quote}
    \small
    \faIcon{comments} \textit{``when it comes to our project manager, we were able to sort of make sure we absorbed some of her the stress that she was under. And to sort of lighten her load.'' - P3 \textcolor{black}{(Developer)}}

\end{quote}

Participants discussed the impact on \textbf{customer management} including \textit{positive impact on customer management} such as customer satisfaction (P3, P5-P7, P10, P14, P16, P19, P22), meeting user and stakeholder expectations (P1, P6), accommodating customer requests (P2, P20), willingness to provide additional product features (P2, P6), and \textit{negative impact on customer management} such as dissatisfied customers (P6, P7, P12, P14, P16, P20). 
Empathy greatly contributed to \textit{customer satisfaction} by ensuring timely, high-quality deliveries and clear communication, which helped developers to align with customer needs, leading to improved satisfaction. 
Participants also described how delivering products that met the technical specifications but failed to meet user expectations or practical needs resulted in \textit{customer dissatisfaction}.
\begin{quote}
    \small
    \faIcon{comments} \textit{``if they know you are doing your best to get things done, and sometimes regardless of your situations, you try to get things done, they know, okay this person is working diligently. And they are also satisfied from the work [..] finally, as a team, you're getting successful. So that increases the client satisfaction.'' - P5 \textcolor{black}{(Developer)}}

\end{quote}

Participants discussed the impact on \textbf{project timelines} including \textit{positive impact on project timelines} such as timely deliveries (P1, P5, P6, P8, P10, P19, P21), and \textit{negative impact on project timelines} such as project delays (P2, P6, P7, P10, P13, P14, P17, P18, P21).
By understanding each other's needs and circumstances, practitioners were able to work efficiently, ensuring \textit{timely deliveries}. 
The absence of empathy led to poor communication and strained relationships, causing \textit{delays in project delivery}.
\begin{quote}
    \small
    \faIcon{comments} \textit{``There were missing deadlines, tasks which were not completed on time. If two weeks of sprint cycle, it usually takes those tasks to complete either three weeks, or four weeks, which were like not properly scoped out.'' - P6 \textcolor{black}{(Developer)}}

\end{quote}

Participants discussed the impact on \textbf{developer efficiency} including \textit{positive impact on developer efficiency} such as increased developer productivity (P1-P3, P5, P6, P9-P11, P13, P17-P19, P22), and \textit{negative impact on developer efficiency} such as decreased productivity (P5, P6, P8, P9, P22).
Empathy \textit{boosted developer productivity} by fostering a positive work environment where team members felt motivated and supported. It enhanced developer knowledge sharing, and communication, leading to increased efficiency and productivity. 
When empathy was lacking, developers felt misunderstood, which \textit{declined their productivity}. Due to this lack of understanding both developers and stakeholders felt unsupported and overwhelmed, resulting in burnout, and further declines in productivity.
\begin{quote}
    \small
    \faIcon{comments} \textit{``being empathic with each other, we are always able to help whenever others face a question. Because the fundamental thing there is, okay I've been there before, so I have the knowledge to help you. So they empathise with each other, therefore they help each other. that definitely speeds up work and there's a lot of productivity gain.'' - P3 \textcolor{black}{(Developer)}}

\end{quote}

Participants discussed the impact on \textbf{resource optimisation} including \textit{positive impact on resource optimisation} such as effective use of resources (P1, P19, P20), and \textit{negative impact on resource optimisation} such as wastage of resources (P1, P14, P19, P20) and increased costs (P6, P18, P19, P20).
Empathy \textit{enabled developers to manage their time effectively} by promoting open communication and clarity around project requirements, which reduced delays and misunderstandings. 
The lack of empathy led to \textit{significant resource wastage}, as developers had to make assumptions due to unclear requirements, resulting in time-consuming mistakes and multiple iterations to meet stakeholder expectations. This extended timelines and wasted effort contributing to \textit{increased project costs}. 
\begin{quote}
    \small
    
    \faIcon{comments} \textit{``Oh costly, wasted our money and everyone's time during that development. because we noticed this project involved different people. Everyone showed their expectations from this software. but without empathy, without communication with each other, to be honest, they spend a lot of money on that and the result was like nothing.'' - P19 \textcolor{black}{(Stakeholder)}}
\end{quote}

\subsection{\textcolor{black}{Negative Impact of Presence of Empathy (Unexpected Consequences of Empathy)}} \label{sec:Negative Impact of Presence of Empathy}
Participants identified several scenarios where the presence of empathy resulted in adverse outcomes, including empathy fatigue (P7, P14, P18, P20), exploitation of empathy (P2, P7, P10), misalignment of priorities (P12, P14, P18), excessive work (P6, P19, P20), negative impact on work-life balance (P10, P21), and increased stress and burnout (P12, P19).
%
Participants experienced instances of \textit{empathy fatigue} when they were overly empathetic towards team members. 
Constant concern for others became emotionally overwhelming, resulting in stress and emotional withdrawal. 
Developers shared instances where their \textit{empathy was exploited}, leading to increased workloads and negatively impacting their productivity and well-being. Highly empathetic individuals often took on more tasks out of a sense of duty or care, which affected their \textit{work-life balance}. Over-committing to tasks, driven by empathy for stakeholders and end-users, led to \textit{increased stress or burnout}.

\begin{quote}
    \small
    \faIcon{comments} \textit{``yes, because it's overwhelming. You have to find a way, like you can't be empathetic to everything all the time, it'll just completely overwhelm you. A lot of tools that we have in project management are there so that you don't have to be empathetic, like they are tools to help you deal with it. So that makes you potentially cold and hard.'' - P20 \textcolor{black}{(Stakeholder)}}
\end{quote}

\subsection{\textcolor{black}{Positive Impact of Lack of Empathy (Unexpected Consequences of Empathy)}} \label{sec:Positive Impact of Lack of Empathy}
Participants identified several scenarios where absence of empathy had positive outcomes, which mostly stemmed from withholding empathy in toxic environments, such as avoiding additional work (P7, P8, P10), timely delivery of projects (P14, P20), direct communication about capabilities \& boundaries (P8, P10), leveraging technical skills of highly skilled developers despite lack of empathy (P12, P18), and work-life balance (P7, P10). 
Developers noted that by not showing empathy, they could \textit{avoid being overwhelmed by additional tasks}. Refusing overtime requests and being direct helped them prevent burnout, maintain a manageable workload, and \textit{establish clear boundaries}, supporting a \textit{healthy work-life balance}. 
Stakeholders acknowledged that some \textit{highly skilled developers, despite lacking empathy}, were still valuable due to their technical expertise. They were able to excel by concentrating on technical tasks without being distracted by emotional factors. 
\begin{quote}
    \small
    \faIcon{comments} \textit{``If I think about developers I've worked with over the years, you know, there are some that have not had high degree of empathy, and really struggled to actually understand other people as human beings. But they could still potentially be very high performing developers and actually code, you know, almost in a silo and isolation.'' - P18 \textcolor{black}{(Stakeholder)}}
\end{quote}

\section{\textcolor{black}{Elaboration of the Contingencies (Strategies)}} \label{sec:Elaboration of the Contingencies}
\textcolor{black}{As introduced in the theory overview (Section \ref{sec:Findings}), we identified several strategies for fostering empathy. In this section, we elaborate on them in detail.}

\subsection{Stakeholder Engagement Related Strategies} \label{sec:Stakeholder Engagement Related Strategies}
This category highlights strategies such as building human-level connections with stakeholders (P2, P5-P9, P16, P22) and promoting empathy through shared experiences (P7, P12). 
Participants noted that developers rarely meet end-users, limiting their empathy. They identified promoting \textit{human-level interactions \textcolor{black}{(Strategy S1)}} between developers and stakeholders as an effective strategy for fostering empathy, 
as direct interactions enhanced understanding of customer needs. 
\textit{Promoting empathy through shared experiences \textcolor{black}{(Strategy S2)}} proved effective. For example, developers observing real users interact with the software allowed them to witness first-hand challenges, fostering empathy and highlighting the need to address user issues.
\begin{quote}
    \small
    \faIcon{comments} \textit{``In a lot of places developer doesn't meet end user, which is pathetic. So a lot of engineers are trained on how to do it. But they don't know why am I building it, and why part is emotionally and empathetically connected. So if you realise that I'm going to build a software for this user, helping them do this, then automatically these things fall in place and you are naturally empathetic towards the larger goal, that I'm going to help this person.'' - P7 \textcolor{black}{(Developer)}}
\end{quote}

\subsection{Communication and Understanding Related Strategies} \label{sec:Communication and Understanding Related Strategies}

Participants shared several \textbf{communication related strategies} including bi-directional communication (P1, P3, P4, P6-P9, P12, P14, P15, P18, P19, P21, P22) and transparency about business goals (P10, P12). 
A \textit{proper bi-directional communication \textcolor{black}{(Strategy S3)}} improved understanding of concerns, and strengthened collaboration. Informal discussions and retrospective meetings further promoted mutual empathy.
\textit{Transparency about business goals \textcolor{black}{(Strategy S4)}} aligned developers' work with organisational objectives, reducing misunderstandings and fostering empathy. 
\begin{quote}
    \small
    \faIcon{comments} \textit{``it's always treating the other person as the expert in their domain, that we are actually hearing them out, having an active communication with them and actually thinking what they might be feeling, when someone might dismiss something and say, oh it should be easy.'' - P18 \textcolor{black}{(Stakeholder)}}
\end{quote}

Participants shared several \textbf{understanding related strategies} including perspective taking (P1, P10, P21, P22), developing a thorough understanding (P1, P7, P9), understanding developers' work (P6, P17, P20), and clarifying doubts (P1, P12). 
\textit{Perspective-taking \textcolor{black}{(Strategy S8)}} led both developers and stakeholders to understand each other's challenges, reducing misunderstandings and enhancing collaboration. \textit{Developing a clear understanding \textcolor{black}{(Strategy S5)}} of team members' roles and pressures, with the help of retrospectives and role-sharing activities fostered empathy and teamwork. Regular meetings helped stakeholders \textit{understand developers' work \textcolor{black}{(Strategy S6)}}, enabling them to offer better support and align expectations.
\textit{Clarifying doubts about user needs and business perspectives \textcolor{black}{(Strategy S7)}} allowed developers to gain insights, leading to better solutions and smoother interactions. 
\begin{quote}
    \small
    \faIcon{comments} \textit{``from my perspective, the key thing for me is understanding what they're doing, what is going on. my understanding of their work helps me in my role, because I have to plan things, and I have to keep stakeholders updated [..] if I know what's going on in the team, then I can convey that to other stakeholders. But if I don't, I can't empathise with what they're doing, then I can't really advocate for the team.'' - P17 \textcolor{black}{(Stakeholder)}}
\end{quote}

\subsection{Team Dynamics and Managerial Approaches} \label{sec:Team Dynamics and Managerial Approaches}
Participants shared \textbf{team dynamics related strategies} including conducting sprint retrospectives (P6, P7, P13, P15, P21), promoting team dynamics (P2, P5, P13), and providing constructive feedback (P13, P17, P18).
\textit{Sprint retrospectives \textcolor{black}{(Strategy S9)}} were instrumental in addressing empathy barriers and enhancing team dynamics. 
\textit{Promoting effective team dynamics \textcolor{black}{(Strategy S10)}} through casual interactions and collaborative problem solving enhanced empathy by fostering personal connections and mutual understanding.
\textit{Providing constructive feedback \textcolor{black}{(Strategy S11)}} encouraged developers to embrace reevaluation positively, and managers guided developers to avoid defensiveness when receiving feedback, framing it as part of system improvement rather than personal criticism. 
\begin{quote}
    \small
    \faIcon{comments} \textit{``when it comes to my past experience, to empathise more with the developers, if you're working with them, if you get to talk with them in personal level, if you know things about his family or personal life, how he is dealing with his life and all, you get to empathise them more.'' - P13 \textcolor{black}{(Stakeholder)}}
\end{quote}

Participants shared \textbf{managerial approaches} such as having backup plans (P10, P17, P20-P22), creating a supportive space (P10, P14, P18, P19), promoting professional \& cross-functional interactions (P1, P11, P12, P17), and strong diversity, equity, and inclusion (DEI) policies (P3).
\textit{Backup plans \textcolor{black}{(Strategy S15)}} provided flexibility in managing resources and timelines during unexpected challenges, demonstrating understanding and support for individual needs while maintaining progress. \textit{Creating a safe and supportive environment \textcolor{black}{(Strategy S14)}} where developers could openly share challenges without fear of judgement also promoted empathy. 
\textit{Promoting professional and cross-functional interactions \textcolor{black}{(Strategy S13)}} helped to bridge communication gaps between technical and business teams, improving understanding and collaboration. 
Interestingly one participant noted that \textit{strong DEI policies \textcolor{black}{(Strategy S12)}} helped to promote empathy by fostering diverse perspectives and inclusive practices, improving collaboration and respect among team members. 
\begin{quote}
    \small
    \faIcon{comments} \textit{``I try having regular one on one meetings, trying to make a safe space where people can say that they don't know something and not felt to be foolish. Try and model that myself, and if I don't know something, I will let it be known that I don't know something. So showing developers that it's okay to not know stuff.'' - P14 \textcolor{black}{(Stakeholder)}}
\end{quote}

\subsection{Training and Education Related Strategies} \label{sec:Training and Education Related Strategies}
Participants shared \textbf{training and education related strategies} including integrating empathy training into IT education \& workplaces (P3, P16, P19), educating IT students on interpersonal skills (P16, P19), using storytelling frameworks (P12, P19), and emphasising the practical benefits of developers' work (P12, P19).
Stakeholders recommended \textit{integrating empathy training into IT education and workplaces \textcolor{black}{(Strategy S16)}}, to help developers become more attuned to the human aspects of their work and enhance their ability to work with customers. 
\textit{Interpersonal skills training \textcolor{black}{(Strategy S17)}} was emphasised as equally important as technical skills. \textit{Storytelling \textcolor{black}{(Strategy S18)}} was identified as a useful tool for fostering empathy, by helping developers to relate to users' challenges. Highlighting the \textit{tangible benefits of developers' work \textcolor{black}{(Strategy S19)}}, how small changes improve user experience, was another effective strategy. 
\begin{quote}
    \small
    \faIcon{comments} \textit{``[..] recommend that everybody working in IT doesn't matter what they're doing, doing IT degree also, does a module on empathy and working with the customer, not against the customer'' - P16 \textcolor{black}{(Stakeholder)}}
\end{quote}

\subsection{Psychological and Attitudinal Approaches} \label{sec:Psychological and Attitudinal Approaches}
Participants shared \textbf{psychological and attitudinal related strategies} including acknowledging developer expertise (P16, P18) and appreciating empathetic behaviour (P7, P20).
%
Stakeholders emphasised \textit{acknowledging developers as experts in their domain \textcolor{black}{(Strategy S20)}}, validated their contributions and helped to maintain team morale and foster a supportive culture. Publicly \textit{acknowledging and appreciating empathetic behaviour \textcolor{black}{(Strategy S21)}} promoted a culture of empathy within teams. 
\begin{quote}
    \small
    \faIcon{comments} \textit{``it's really about where you and developer meet. it's about, the empathy at the point of interpretation of the brief. And acknowledgement that not all clients are silly, and it's about suspending judgement on both sides.'' - P16 \textcolor{black}{(Stakeholder)}}
\end{quote}

\textcolor{black}{An overview of the strategies discussed above is provided in Table \ref{tab:Empathy Contingencies}}.

\begin{table*}[htbp]
    \centering
    \scriptsize
    \caption{Empathy Contingencies \textcolor{black}{(Strategies)}}
    \label{tab:Empathy Contingencies}
    \begin{tabular}{P{0.03\textwidth} P{0.4\textwidth} P{0.03\textwidth} P{0.4\textwidth}}%
        \toprule
        \textbf{ID} & \textbf{Empathy Contingencies \textcolor{black}{(Strategies)}} & \textbf{ID} & \textbf{Empathy Contingencies \textcolor{black}{(Strategies)}}\\
        \midrule

        S1 & Fostering human level interactions with stakeholders & S12 &  Having strong DEI policy\\
        S2 & Promoting empathy through shared experiences & S13 & Promoting professional and cross-functional interactions\\
        S3 & Having proper bi-directional communication & S14 & Creating a safe and supportive space for developers\\
        S4 & Transparency about business goals & S15 & Having backup plans \\
        S5 & Developing a thorough understanding & S16 & Integrating empathy training\\
        S6 & Understanding developers' work & S17 & Educating IT students on interpersonal skills\\
        S7 & Clarifying doubts & S18 & Using storytelling frameworks to facilitate understanding\\
        S8 & Perspective taking & S19 & Emphasising the practical benefits of developers' work\\
        S9 & Conducting the sprint retrospectives & S20 & Acknowledging developer expertise\\
        S10 & Promoting effective team dynamics & S21 & Appreciating empathetic behaviour\\
        S11 & Providing respectful and constructive feedback to developers & &\\
        \bottomrule
     \end{tabular}
\end{table*}

\section{Discussion} \label{sec:Discussion}

\textcolor{black}{While Section \ref{sec:Findings} is focused on key findings, this section delves into their broader implications within the context of SE. Section \ref{sec:Discussion of findings in relation to existing literature} offers a more detailed analysis by relating the key findings to existing literature. 
We also examine the implications of our findings for future research (Section \ref{sec:Implications for Research}) and for practical applications in SE (Section \ref{sec:Implications for Practice}).}

\subsection{Positioning Our Findings Within Existing Research} \label{sec:Discussion of findings in relation to existing literature}


\textcolor{black}{While familiarity \cite{motomura2015interaction, airenti2015cognitive}, relatedness \cite{jami2023interaction, airenti2015cognitive}, and culture \cite{jami2023interaction, halabi2008social, kitayama2000culture, schwartz2010biculturalism, chopik2017differences} are well-established causes of empathy, studying empathy in SE offers unique insights. Technical complexity, interdisciplinary collaboration, and distributed teams of SE introduces challenges that shape how empathy manifests, making it crucial to understand these dynamics within this domain. Our findings confirm these general causes while identifying SE-specific factors, such as the role of technical expertise and work performance in empathy development (Section \ref{sec:Causes of Empathy (Enablers)}). Similarly the general causes of a lack of empathy such as time pressure, negative emotions, and task-centeredness are well-known in other fields \cite{derksen2016managing, howick2017barriers, halpern2003clinical, taleghani2018barriers}. SE presents role-based dynamics, such as the conflicting goals of developers and testers, which exacerbate challenges that are less pronounced in other domains (Section \ref{sec:Causes of Lack of Empathy (Barriers)}). Our study identifies SE specific inhibitors, including organisational and environmental gaps, demonstrating that while some empathy barriers overlap with other domains, their manifestation in SE presents distinct challenges and novel insights (Section \ref{sec:Causes}).}

\textcolor{black}{Empathy improves well-being \cite{hojat2016empathy, decety2015empathy, derksen2013effectiveness, decety2020empathy, moudatsou2020role, de2022impact}, quality \cite{hess2016voices, strobel2013empathy}, collaboration \cite{hess2016voices, strobel2013empathy, min2015factors}, trust and satisfaction \cite{aggarwal2005salesperson, min2015factors} across various fields, including healthcare, engineering, and sales. Similarly, in SE, empathy improves practitioner well-being, trust, product quality, and collaboration \cite{cerqueira2023thematic, cerqueira2024empathy, lind2024empathy}. Our study corroborates these effects in SE (Section \ref{sec:Impact on mental health and well-being of practitioners}). In addition, our study found that empathy influences technical outcomes like software quality, code maintainability, and user-centred design, with empathetic developers producing more intuitive software and reducing bugs (Section \ref{sec:Impact on Software Development practices}). This direct link between empathy and technical outcomes is unique to SE and represents a key contribution. While positive effects of empathy on well-being are known (Section \ref{sec:Impact on mental health and well-being of practitioners}), the specific demands of SE such as long hours, tight deadlines, and rapid iteration cycles amplify the importance of empathy in team cohesion and reducing burnout. By investigating these SE-specific effects, our study confirms the general benefits of empathy while showing its direct influence on software project success, and productivity.}

\textcolor{black}{While prior research has developed empathy theories from a psychological perspective (Section \ref{sec:Related Works}), our theory introduces a comprehensive SE-specific framework addressing both interpersonal and technical aspects of empathy. Existing models \cite{burns2021theodor, hoffman1996empathy, wondra2015appraisal, gunatilake2023empathy} overlook the influence of SE methodologies, technical constraints, and cognitive demands. To the best of our knowledge, our theory is the first to provide an in-depth exploration of empathy in SE, showing how empathy shapes technical complexity and development workflows. Unlike traditional models, it emphasises the evolving, situational nature of empathy within software teams, shaped by organisational culture, stakeholder interactions, and team dynamics. While existing research acknowledges benefits of empathy in SE \textcolor{black}{\cite{cerqueira2023thematic, cerqueira2024empathy, lind2024empathy}}, our work offers a structured understanding of empathy barriers and actionable strategies to address them. With insights from diverse SE roles, our theory provides a broader perspective on the role of empathy in SE that not only builds on existing empathy literature but also offers practical recommendations to enhance empathy, improving processes and outcomes in SE.}

\subsection{Implications for Research} \label{sec:Implications for Research}
\textcolor{black}{This section outlines key implications for future research, directly informed by our findings. Each insight is grounded in the data, with clear connections to our findings. These implications offer a structured pathway for advancing research and ensure their relevance within the broader context of empathy in SE.}

\noindent \faIcon{graduation-cap} \textbf{Research on factors influencing empathy:} We found several factors influencing empathy including culture, personality, job role, and gender (Section \ref{sec:Conditions}). 
Further research is needed to uncover any additional factors that may influence empathy and assess whether their impact is positive or negative. 

\noindent \faIcon{graduation-cap} \textbf{Interplay of different roles in relation to empathy:} Interactions between SE roles vary based on the nature of each role (Section \ref{sec:Causes of Empathy (Enablers)}, \ref{sec:Conditions}). For instance, developers and testers often experience friction, whereas testers and product owners tend to have more harmonious interactions. Empathy expressed among these groups depends on these dynamics. While this study examined these relationships collectively, further in-depth research into role-specific interactions could offer more nuanced insights into how empathy is influenced and expressed in different SE roles.

\noindent \faIcon{graduation-cap} \textbf{Role of leadership in fostering an empathetic organisation and team culture:} Our findings show that teams tend to be more empathetic when led by an empathetic manager or team lead, indicating that empathy is influenced by leadership (Section \ref{sec:Causes of Empathy (Enablers)}, \ref{sec:Impact on mental health and well-being of practitioners}). Further research could offer deeper insights into how leadership influences the development and manifestation of empathy within teams and organisations. 

\noindent \faIcon{graduation-cap} \textbf{Empathy training frameworks:} Empathy training is widely used in medical education to enhance patient-clinician empathy, with studies demonstrating its effectiveness \cite{gunatilake2023empathy, teding2016efficacy}. Given the positive role of empathy in SE, developing and implementing empathy training for software practitioners could be highly beneficial \textcolor{black}{(P3, P11-P14, P16, P18, P19, P20)}. Research into effective frameworks and methods to assess the longevity of the benefits of such trainings would enhance the effectiveness and sustainability of these programs.

\noindent \faIcon{graduation-cap} \textbf{Assessing the impact of empathy:} We examined the nuanced effects of empathy on both technical and human aspects of SE. However, there is no method to systematically assess or validate these impacts \textcolor{black}{\cite{gunatilake2023empathy}}. Developing a SE-oriented empathy scale would address this gap, providing a tool to effectively measure and evaluate the influence of empathy in SE. Without such a scale, objectively assessing the influence of empathy in SE remains difficult.

\noindent \faIcon{graduation-cap} \textbf{More research on the surprising effects of empathy:} We identified both negative effects of presence of empathy, such as empathy fatigue, and positive effects of its absence (Section \ref{sec:Negative Impact of Presence of Empathy}, \ref{sec:Positive Impact of Lack of Empathy}). 
Studying the negative impacts, such as empathy fatigue, could provide valuable insights into applying empathy effectively and selectively. Similarly, exploring the positive effects of the absence of empathy could help identify situations where empathy may be counterproductive. Understanding when empathy should or should not be applied would optimise its use in SE practices.

\subsection{\textcolor{black}{Implications for Practice}} \label{sec:Implications for Practice}

\textcolor{black}{\noindent \faIcon{laptop} \textbf{Bridging silos through empathy:} Empathy plays a critical role in narrowing the divide between technical and non-technical stakeholders, improving collaboration, and aligning software development with broader business needs (Section \ref{sec:Impact on Software Development practices}). Participants highlighted that misunderstandings often stem from communication gaps and limited exposure to each other’s contexts. Empathetic communication through clear, two-way dialogue, tailored explanations, and active listening helps practitioners appreciate diverse roles and challenges. Transparency about business goals and involving both technical and non-technical practitioners in decision making were also seen to promote shared understanding and trust. Strengthening these practices can break down role-based silos, enhance interpersonal dynamics, and ultimately lead to better software outcomes.}

\textcolor{black}{\noindent \faIcon{laptop} \textbf{Empathy in agile ceremonies:} Embedding empathy into agile ceremonies can improve collaboration and enhance overall development outcomes (Section \ref{sec:Impact on Software Development practices}, Section \ref{sec:Team Dynamics and Managerial Approaches}). During sprint planning, empathy enables realistic workload distribution by acknowledging individual capacities and constraints. In stand-ups, it facilitates open communication and timely support for emerging challenges. Empathetic retrospectives and reviews foster psychological safety, encourage honest reflection, and strengthen team cohesion. By modelling empathy in these routine practices, both practitioners and leaders can cultivate a more supportive team culture.}

\textcolor{black}{\noindent \faIcon{laptop} \textbf{Well-being and retention:} Organisations that prioritise empathy in their culture and leadership can create healthier, more inclusive work environments, which can help reduce employee burnout and improve retention (Section \ref{fig:Impact on mental health}). Empathetic organisations are more likely to offer support systems for employees dealing with stress, workload challenges, or personal issues, creating a workplace where individuals feel valued and cared for. When employees feel that their well-being is prioritised, they are more likely to remain engaged, motivated, and loyal to the company. Further, fostering empathy within teams can lead to stronger interpersonal relationships and a sense of belonging, which are crucial for overall job satisfaction. By creating an empathetic work environment, organisations can not only improve employee mental health but also reduce turnover, leading to a more stable and productive workforce.}

\textcolor{black}{\noindent \faIcon{laptop} \textbf{Managing empathy fatigue:} While empathy is essential for fostering positive relationships and collaboration, excessive emotional engagement can lead to empathy fatigue (Section \ref{sec:Negative Impact of Presence of Empathy}). Participants in our study highlighted the importance of setting clear boundaries and prioritising self-care to sustain emotional well-being. Normalising conversations around workload, emotional strain, and the need for rest can help reduce stigma and prevent burnout. Organisations that support mental health through accessible resources such as counselling services or flexible work policies can create more resilient teams. Acknowledging and addressing empathy fatigue is vital for maintaining a healthy and sustainable work environment, particularly in high-pressure or emotionally demanding projects.}

\textcolor{black}{\noindent \faIcon{laptop} \textbf{Empathy education and training:} Integrating empathy and interpersonal skills into SE education and workplace training can enhance collaboration and emotional intelligence within teams (Section \ref{sec:Training and Education Related Strategies}). Participants in our study emphasised the importance of cultivating empathy early, noting that SE curricula should include dedicated modules on communication and relationship-building. In industry settings, empathy training can support practitioners in applying these skills in real-world scenarios, improving their ability to navigate diverse team dynamics and stakeholder interactions. Promoting empathy through structured learning initiatives helps foster respectful, inclusive, and emotionally aware work environments, contributing to both improved software practices and practitioner well-being.}

\section{Evaluation} \label{sec:Evaluation}
Evaluating STGT involves evaluating both research method application and the outcomes of the study \cite{hoda2022STGT, hoda2024qualitative}. These aspects are detailed in the following Section \ref{sec:Evaluating STGT application} and \ref{sec:Evaluating the theory outcome}.

\subsection{Evaluating STGT application} \label{sec:Evaluating STGT application}
The two criteria for evaluating the application of STGT are \textit{credibility} and \textit{rigour} \cite{hoda2022STGT, hoda2024qualitative}. Credibility ensures that the research was conducted in accordance with established guidelines, while rigour ensures that the research was carried out to a high standard. Meeting these criteria demonstrates the quality of the STGT application.
We establish \textit{credibility} by providing detailed descriptions of the data collection and data analysis phases in Section \ref{sec:Research Method}. This includes information on participant recruitment, sampling techniques, the iterative and interleaved nature of data collection and analysis, application of theoretical sampling, mode of theory development, interview guide refinements, theoretical saturation, theoretical integration, memos, applied research paradigm, and the rationale for all the methodological decisions. 
Evidence of \textit{rigour} is demonstrated by providing examples of sanitised raw data (interview quotes) throughout Section \ref{sec:Findings}. We illustrate how coding procedures such as open coding and constant comparison were applied to derive codes, concepts, categories (see Figure \ref{fig:STGT Example}). Further evidence of the advanced stages of theory development is provided in Section \ref{sec:Research Method}, where we detail the use of memos, targeted coding, and the process of theoretical integration.

\subsection{Evaluating the Theory Outcome} \label{sec:Evaluating the theory outcome}
An outcome of the full STGT method is considered mature when it consists of well-supported categories and relationships. Mature theory outcomes must be evaluated using higher standards than preliminary ones, demonstrating characteristics such as novelty, usefulness, parsimony, and modifiability \cite{hoda2022STGT, hoda2024qualitative}. Our paper presents a \textit{novel} theory on the role of empathy in developer-stakeholder interactions in SE. \textcolor{black}{As discussed in Section \ref{sec:Related Works}, empirical research on empathy in SE remains limited, and to the best of our knowledge, this is the first theory that systematically examines empathy within SE practice. In addition, we review existing empathy theories and models in Section \ref{sec:Related Works} and directly compare our theory with them in Section \ref{sec:Discussion of findings in relation to existing literature}, highlighting its distinct contributions to the field. Unlike existing psychological theories of empathy, which primarily focus on different conceptualisations of empathy, our theory specifically addresses the socio-technical nature of the role of empathy as it plays out in developer-stakeholder interactions in an SE context. It accounts for the unique challenges posed by SE, such as distributed teams, technical constraints, and agile methodologies, which shape how empathy is experienced and expressed in SE contexts.}
Our theory is \textit{useful} for software practitioners (domain actors), as it can improve their interactions, mental health and well-being, while enhancing software development practices. It offers recommendations for practice and recommendations for future research. 
Our theory demonstrates \textit{parsimony} by structuring complex theoretical outcomes through the 6Cs template, capturing key categories and relationships in a compact and succinct way, while ensuring that concepts and categories remain cohesive and integrated. Our theory is \textit{modifiable}, allowing for extension with new findings or additional depth, such as extending to different contexts by exploring specific interactions between SE groups, \textcolor{black}{including} developers and testers, or testers and product owners.

\section{Limitations} \label{sec:Limitations}
\textcolor{black}{While our study provides valuable insights into the role of empathy in SE, it is important to acknowledge certain limitations. To structure this discussion, we adopted the Total Quality Framework (TQF) for qualitative research \cite{roller2015tqf}, which is more suited to our study than traditional validity frameworks encompassing internal, external, construct, and conclusion validity, typically used in quantitative research \cite{lenberg2024qualitative}. Below, we outline key limitations across four components of TQF: \textit{credibility} (related to data collection), \textit{analysability} (focused on data analysis), \textit{transparency} (regarding reporting), and \textit{usefulness} (concerning the applicability of findings).}

\textcolor{black}{The \textit{credibility} component concerns the completeness and accuracy of data collection \cite{roller2015tqf} and consists of two elements: scope and data gathering.  
A key limitation related to the \textit{scope} element is the generalisability of findings, constrained by our sample composition, which, though diverse, did not explore specific roles or domains in-depth. As with most qualitative studies, our findings are limited to the contexts we studied, which are further constrained by the participants we were able to access. To enhance representation, we recruited participants from various countries, considering gender, geography, and roles. However, future research could further expand the participant pool to explore aspects of our theory in greater depth and breadth. 
While our study reached theoretical saturation, grounded theories remain adaptable, allowing future research to refine or expand key categories in different contexts \cite{hoda2022STGT, hoda2024qualitative, glaser2017discovery}. Thus, while our findings are grounded in participant experiences, the developed theory may evolve with future studies.}
\textcolor{black}{A key limitation of \textit{data gathering} is the subjective interpretation of empathy, which can vary across individuals. To address this, we asked participants to define empathy both generally and in SE contexts, using scenario-based questions to align their perspectives with their experiences. While this approach helped to capture practitioners' authentic lived experiences rather than imposing a theoretical definition that might not resonate with their diverse contexts, their interpretations may still vary. In addition, categorising participants as ``human-centric'' or ``technology-centric'' was based on self-assessment of the participants, which may introduce subjectivity.}

\textcolor{black}{The \textit{analysability} component addresses the completeness and accuracy of data analysis \cite{roller2015tqf} involving two elements: processing and verification. A potential limitation is the subjectivity in data interpretation. While we followed a rigorous and iterative data analysis using STGT, researcher interpretations may still influence the findings. To minimise this, the first author’s initial codes were reviewed by co-authors, and the third author, an STGT expert, peer-reviewed all elements. After multiple review rounds, the codebook was collaboratively finalised. In addition, participant self-reports on empathy related experiences may be influenced by social desirability bias, where individuals may overstate or understate their empathetic behaviours.}
\textcolor{black}{The imbalance between human-centric and technology-centric perspectives in the sample may have influenced the findings. Sixteen out of 22 participants were more human-centric, which could have led to an emphasis on empathy in the findings. Future studies with a more balanced sample would offer a broader view. While STGT’s constant comparison approach ensured concept validity, the lack of statistical sampling means additional participants may reveal new insights. Theoretical saturation was reached after interviews with participant P20, and further interviews confirmed no new insights, ensuring the developed theory is well-supported.}
\textcolor{black}{We also acknowledge the ethical and interpretive challenges involved in presenting participants’ views on personality traits as potential influences on empathy. While some participants associated these traits (e.g., introversion) with reduced empathetic engagement, we recognise the risk of reinforcing stereotypes or stigmatising certain temperaments. For example, equating introversion with a lack of empathy may oversimplify a complex interplay of individual disposition, context, and interactional norms. As researchers, we have made conscious efforts to present such perspectives as participant perceptions, not empirical generalisations or normative claims. We have also critically examined the language used in reporting these concepts to avoid circular reasoning and stereotypical trait assumptions. Despite these efforts, we acknowledge that unconscious bias may still influence how participant views are interpreted and framed. Future work could further explore how empathy is shaped by diverse personality traits across varying social and organisational contexts, while maintaining ethical sensitivity and reflexivity.}

\textcolor{black}{The \textit{transparency} component of TQF emphasises the importance of clear and comprehensive documentation, covering all aspects related to credibility and analysability. This is achieved through thick description, which includes rich details to help readers assess the transferability of the study’s methods, findings, and recommendations. We have thoroughly explained the research design (Section \ref{sec:Research Method}), findings (Section \ref{sec:Findings}), and implications (Sections \ref{sec:Implications for Research} and \ref{sec:Implications for Practice}). This ensures that readers can critically evaluate and apply our methods and findings in different contexts.}

\textcolor{black}{The \textit{usefulness} component focuses on the practical value of the study outcomes. Our findings offer value to both SE researchers, encouraging further exploration in this area, and to software practitioners, helping them improve mental health and wellbeing while enhancing software development practices. Further, to the best of our knowledge, no prior theories have explored empathy in SE, making our work a novel contribution to the field.}
\textcolor{black}{While this study offers valuable insights into empathy in SE, the limitations discussed may impact the interpretation and generalisation of our findings. Despite these limitations, our study provides valuable insights and a foundation for future research.}

\section{Conclusion} \label{sec:Conclusion}
We conducted a socio-technical grounded theory (STGT) study involving 22 semi-structured interviews with software developers and their stakeholders to understand the role of empathy in SE focusing on the interactions between these groups. Using STGT, we analysed the data and developed a comprehensive theory of empathy in SE. Framed by the 6Cs template, our theory address the context, conditions, causes, and consequences of empathy, as well as its absence. We also identified contingencies for enhancing empathy and explored covariances between these categories. Our findings highlight that empathy is a key factor influencing both the personal well-being of software practitioners and the success of software development processes. Based on our findings, we offer practical implications for software practitioners and propose future research directions for the academic community to further explore the role of empathy in SE.

\begin{acks}
Gunatilake and Grundy are supported by ARC Laureate Fellowship FL190100035. We express our gratitude to all our interview participants for generously sharing their lived experiences, which grounded our research in real-world insights; this work would not have been possible without their contributions.
\end{acks}

\bibliographystyle{ACM-Reference-Format}
\bibliography{sample-base}

\end{document}